\documentclass{aastex}
\usepackage{emulateapj5}
\usepackage{color}
\usepackage{onecolfloat}
\usepackage{epsf}
\usepackage{ulem}
\newcommand{\erf}{\rm erf}

\newcommand{\Ml}{M_{\rm 1}}
\newcommand{\Mm}{M_{\rm min}}
\newcommand{\Mlft}{M_{\rm 1}}
\newcommand{\Mmft}{M_{\rm min}}

\newcommand{\fcen}{f_{\rm cen}}

\newcommand{\nfo}{${\rm n_4}$}
\newcommand{\zoM}{Models 1--11}
\newcommand{\ztM}{Models 21--31}
\newcommand{\zHzoM}{Models 1H--11H}
\newcommand{\zHztM}{Models 21H--31H}
\newcommand{\zoMfn}{Models 7${\rm n_4}$--10${\rm n_4}$}
\newcommand{\al}{\alpha}
\newcommand{\Om}{\Omega_m}
\newcommand{\Ob}{\Omega_b}

\newcommand{\OL}{\Omega_\Lambda}

\newcommand{\hMpc}{h^{-1}{\rm\;Mpc}}

\newcommand{\itrihMpc}{h^{3}{\rm\;Mpc^{-3}}}

\newcommand{\Vbox}{V_{\rm box}}

\newcommand{\ncen}{N_{\rm cen}}
\newcommand{\nsat}{N_{\rm sat}}

\newcommand{\Msun}{M_{\odot}}
\newcommand{\mng}{\langle N_{\rm g}{\rm (M)} \rangle}
\newcommand{\mngc}{\langle N_{\rm cen}{\rm (M)} \rangle}
\newcommand{\mngs}{\langle N_{\rm sat}{\rm (M)} \rangle}

\newcommand{\lan}{\langle}
\newcommand{\ran}{\rangle}
\newcommand{\NNs}{\langle N_{\rm g}(N_{\rm g}-1) \rangle_M}
\newcommand{\NNsN}{\langle N_{\rm g}(N_{\rm g}-1) \rangle /\langle N_{\rm g} \rangle^2}

\newcommand{\NNss}{\langle N_{\rm sat}(N_{\rm sat}-1) \rangle_M}

\shorttitle{PASSIVE EVOLUTION OF GALAXY CLUSTERING}
\shortauthors{SEO, EISENSTEIN, \& ZEHAVI}

\begin{document}
\twocolumn[%
\submitted{Accepted for publication in \textit{The Astrophysical Journal}}
\title{
Passive Evolution of Galaxy Clustering}
\author{Hee-Jong Seo\altaffilmark{1,2},
Daniel J.\ Eisenstein\altaffilmark{1},
Idit Zehavi\altaffilmark{3} }
\email{sheejong@fnal.gov,deisenstein@as.arizona.edu,izehavi@abacus.astr.cwru.edu}

\begin{abstract}
We present a numerical study of the evolution of galaxy clustering when galaxies flow passively from high redshift, respecting the continuity equation throughout. While passive flow is a special case of galaxy evolution, it allows a well-defined study of galaxy ancestry and serves as an interesting limit to be compared to non-passive cases. We use dissipationless N-body simulations, assign galaxies to massive halos at $z=1$ and $z=2$ using various halo occupation distribution (HOD) models, and trace these galaxy particles to lower redshift while conserving their number. We find that passive flow results in an asymptotic convergence at low redshift in the HOD and in galaxy clustering on scales above $\sim 3\hMpc$ for a wide range of initial HODs. As galaxies become less biased with respect to mass asymptotically with time, the HOD parameters evolve such that $\Ml/\Mm$ decreases while $\al$ converges toward unity, where $\Mm$ is the characteristic halo mass to host a central galaxy, $\Ml$ is the halo mass to host one satellite galaxy, and $\al$ is the power-law index in the halo-mass dependence of the average number of satellites per halo. The satellite populations converge toward the Poisson distribution at low redshift. The convergence is robust for different number densities and is enhanced when galaxies evolve from higher redshift. We compare our results with the observed Luminous Red Galaxy (LRG) sample from Sloan Digital Sky Survey that has the same number density. We claim that if LRGs have experienced a strict passive flow, their $\mng$ should be close to a power law with an index of unity in halo mass. Discrepancies could be due to dry galaxy merging or new members arising between the initial and the final redshifts.
The spatial distribution of passively flowing galaxies within halos appears on average more concentrated than the halo mass profile at low redshift.
The evolution of bias for passively flowing galaxies is consistent with linear bias evolution on quasilinear as well as large scales. 
\end{abstract}

\keywords{galaxy clustering
--- passive flow evolution
--- halo occupation distribution
--- methods: N-body simulations
}
]

\altaffiltext{1}{Steward Observatory, University of Arizona,
                933 N. Cherry Ave., Tucson, AZ 85121}

\altaffiltext{2}{Center for Particle Astrophysics, Fermi National Accelerator Laboratory, P.O. Box 500, Batavia, IL 60510-5011}
\altaffiltext{3}{Department of Astronomy, Case Western Reserve University, 10900 Euclid Ave, Cleveland, Ohio 44106}

\section{Introduction}
An accurate match between galaxies at different redshifts can provide essential clues for constructing theories of galaxy evolution. The ancestry of a given population of galaxies, however, is complicated to establish: as galaxies undergo major or minor merging with different probabilities or have different star-formation rates, individual galaxies may no longer retain the common internal, observational properties at later redshifts. Meanwhile, the evolution of {\it clustering} of the given initial galaxy population as a whole can provide another route to trace ancestry, in principle, independent of the evolution of the appearance (i.e., the internal properties) of galaxies: except for on very small scales where the baryonic physics may dominate, the clustering of galaxies depends on the clustering of the typical underlying density peaks where they reside. Understanding the evolution of clustering of galaxy populations can provide a key tool for controlling the uncertainties in the evolution of the appearance of galaxies. As a starting point, in this paper we will study how a given galaxy population is spatially distributed at low redshift under the assumption that a single unchanging set of galaxies flows from various initial distributions at high redshift. In other words, we consider the effect of the continuity equation in gravitational clustering, with no sources or sink terms due to galaxy merging or formation.

Recent studies show that one can reproduce the clustering of observed galaxies by taking the halos and their subhalos as a proxy between the galaxies and mass with a proper correspondence of galaxy luminosity to halo mass \citep{Colin99,Neyrinck04,Krav04,Conroy06,Weinberg06}. The resulting occupancy of the galaxies among halos is often parameterized in the form of halo occupation distribution at the given redshift (hereafter, HOD) \citep{Benson00,Peacock00,Seljak00,Sco01,Benson01a,White01,Berlind02,Berlind03,Krav04,Zehavi05b,Zheng05,ZW07,Zheng07a}. 
In those semi-analytic and SPH simulations, galaxy populations are selected by stellar or baryon mass that would correspond to observational properties at the given time. Such snapshots at different redshifts collectively offer a general idea on the evolution of clustering and halo occupation statistics of galaxies. For example, the distinction between old and young populations in \citet{Berlind03} and \citet{Zheng05} hints an evolution of HOD of galaxies with time. 

However, the characteristics of the clustering evolution of a given population are difficult to directly infer from these studies, as here the set of galaxies at low redshift does not necessarily have a one-to-one relation with the set at high redshift. Conversely, the connection between populations at different epochs can be tested and confirmed once we understand the characteristics of the clustering evolution for a given population by means of N-body simulations.

In addition, once we understand the evolution of clustering of galaxies, we also acquire a better handle for the clustering bias between the galaxies and mass, which is crucial for deriving cosmological information from the galaxy distribution. That is, to isolate useful cosmological information from the observed galaxy distributions, we need a good theory to predict the characteristics of galaxy correlation function for a given population so that we can reduce the uncertainty in modeling galaxy clustering. 

What is meant by `a given population of galaxies' of course depends on our definition. By defining a population at the initial redshift regardless of their appearance at the observed time, we naturally exclude the addition of any new members to the population during evolution. The galaxies of a given initial population in reality may undergo merging among them, in which case members of the population at different redshifts no longer have a one-to-one correspondence. In this study, we want to test the clustering evolution of a galaxy population when its members are preserved throughout evolution. As such, the population respects the continuity equation without source (i.e., new members) and sink terms (i.e., merging) and therefore its number is conserved. We refer to this as `passive flow' evolution of a galaxy population, as an analogy to the passive stellar evolution where the stellar population within an individual galaxy is preserved. Although this passive flow is certainly not the whole picture of galaxy evolution, it serves as an interesting limit of galaxy clustering evolution by itself and so as a template to be compared to non-passive flow evolution. It is shown that the continuity equation naturally forces the evolution of clustering of biased tracers to converge to a small area of parameter space, at least on large scales: the bias on large scales converges toward unity with time {\citep{Dekel87,Nusser94,Fry96,Hamilton97,Tegmark98}}. In this paper we test whether the continuity condition, i.e., passive flow, results in further distinct parametric signatures.

We use dissipationless N-body simulations, identify halos, and assign galaxies to halos based on various halo occupation distributions. We trace these galaxies to lower redshifts while conserving their numbers, and study the evolution of HOD and the correlation function in the turn-around regime (i.e., quasilinear or smaller scales). 

We then compare the results of passive flow with observed galaxy populations. While galaxies flow passively, the stellar contents of individual galaxies may or may not passively evolve. Even without merging, some galaxies within a given initial population may continue to form stars with existing cold gas and stay blue, while others might end star formation and become red. However, such a heterogeneous appearance makes it difficult to select a consistent set of objects at the two redshifts, i.e., to avoid sources and sink terms. We thus consider red galaxies that have lost their cold gas content and ceased star formation before $z\sim 1$ or $2$ (i.e., galaxies with passive stellar evolution) as an observational counterpart. Such galaxies will likely maintain a common observational appearance during the subsequent evolution. When comparing to the observations, we will look for any discrepancies in galaxy correlation function and HOD that might indicate galaxies entering or leaving the observed population. Hereafter, we often use the term galaxy `evolution' to describe the evolution of galaxy distributions through passive flow. We will explicitly use `passive stellar evolution' to distinguish from passive flow. An example of such comparison between the numerical passive flow evolution and the observed red galaxies is found in \citet{White07}. In this paper, we aim at a more extensive study of characteristics of passive flow evolution.

We compare our results with Luminous Red Galaxies (hereafter, LRGs) from Sloan Digital Sky Survey (hereafter, SDSS) at $z=0.3$ \citep{Eis01}, as an observational counterpart. These galaxies reside in very massive halos, and their red colors, lack of cold gas, and the evolution of luminosity function imply that they have had low star formation rates for a few billions of years \citep[e.g.,][]{Wake06,Brown07}. The observed small-scale clustering implies their low merger rate, at least at low redshift \citep{Mas06}. Therefore the comparison of our results to the LRGs will manifest signatures of non-passive flow evolution in the LRGs if any. With these realistic counterparts available, we focus our study of passive flow evolution on galaxies that give the LRG number density at low redshift. We call these galaxies as `progenitors' of the LRGs. We do not assume any specific observational properties of these progenitors, but only assume that these progenitors are a population that is distributed in halos {\it as a function of the host halo mass at the initial redshift} and {\it are set to undergo passive flow after the initial redshift}. We look for any dependence of the outcome on the initial redshift. We also study the passive flow evolution for lower mass halos to find whether the signatures of passive flow depend on a halo mass range. 

In \S~\ref{sec:nbody} we describe the parameters for N-body simulations and methods for assigning LRG progenitors at high redshift. In \S~\ref{sec:z1Mz2M} we present the results of the first moments of HOD, the average number of galaxies per halo as a function of halo mass, and correlation function of passively flowing galaxies. In \S~\ref{sec:z1Mz2MNN} we investigate the second moments of HOD, the average pair counts within a given mass of halo, for passively flowing galaxies. In  \S~\ref{sec:z0.3M} we compare the resulting statistics for passively flowing galaxies with the best fit for the observed LRGs, and in \S~\ref{sec:z0.3H} we search for a signature of passive flow evolution or environmental effects in correlation function for the given HOD. In \S~\ref{sec:BiasRed} we study the evolution of bias, compared to linear theory. In \S~\ref{sec:evolCor} we discuss the details in the clustering of passively flowing galaxies.

\section{Simulated halos and galaxy populations }\label{sec:nbody}
\subsection{N-body simulations}
Our cosmological N-body simulations use the Hydra code \citep{couch95} in collisionless ${\rm AP^3M}$ mode. We use the CMBfast \citep{SeZa96,Za98,ZaSe00} linear power spectrum to generate many initial Gaussian random density fields at redshift of 49 and evolve them to lower redshifts. We generate the initial fields using the cosmological parameters similar to the 1st year Wilkinson Microwave Anisotropy Probe (WMAP) data \citep{Spergel03}: $\Om=0.27$, $\OL=0.73$, $\Ob=0.046$, $h=0.72$, and $n=0.99$. We normalize the initial fields by requiring $\sigma_8 = 0.9$ at $z=0$ and assuming a linear growth function. 
Each simulation represents $\Vbox=256^3h^{-3}{\rm\;Mpc^3}$ and follows the evolution of $256^3$ dark matter particles ($\sim 1.0355 \times 10^{11} \Msun$/particle). We compute gravity using $256^3$ force grids with a Plummer softening length of 0.1 $\hMpc$ in comoving unit. A total of 29 simulations are used to allow little interference from statistical variance. 

We use the friends-of-friends method \citep{Davis85} and identify host halos by adopting a comoving linking length of 0.2 $\hMpc$. We assign galaxies to dark matter halos with various halo occupation models.  In our models, halos with fewer than 100 particles host at most $0.2\%$ of the galaxy population at $z=1$ and $3\%$ at $z=2$. We use small group multiplicities only to represent the extreme low mass tail of the halo occupation distribution.

\subsection{HOD models and galaxies} \label{subsec:models}
We start with an assertion that the number of galaxies in a halo at the initial redshift is only a function of the halo mass not of environments. We assume that the initial halo occupation distribution follows those observed in the local universe and adopt the following form:

\begin{equation}\label{eq:Ngal}
\mng = \exp(-\Mm/M) \times \left[ 1+(M/\Ml)^\al \right]
\end{equation}

\noindent where $M$ is the mass of a halo, $\Mm$ is a characteristic mass scale for a halo to have one central galaxy, and $\Ml$ is a mass for a halo with a central galaxy to have one satellite galaxy. A central galaxy is assigned to a halo based on the nearest integer distribution with the average of $\lan \ncen (M) \ran = \exp(-\Mm/M)$. We find the most bound particle\footnote{We find a particle with the lowest total energy in a halo.} within each halo and label it as a central galaxy. For those halos hosting central galaxies, a number of satellites are randomly assigned to the rest of particles in the halo, based on the Poisson distribution with the average of $\lan \nsat (M) \ran=(M/\Ml)^\al$ (that is, the average of $\lan \nsat (M) \ran=\exp(-\Mm/M)(M/\Ml)^\al $ over all halos). The satellite galaxies therefore trace mass inside the halo at the initial redshift.

The observed spatial distribution of galaxies within a halo is similar to mass distribution while the subhalo distribution from simulations is antibiased with respect to mass and galaxies \citep[][ and observational references therein]{Ghigna98,Ghigna00,Colin99,Springel01,Diemand04,Gao04a,Gao04b,vanB05a,Nagai05,Zentner05,Weinberg06} \citep[but see][]{Taylor04}. Meanwhile, \citet{Nagai05} and \citet{Conroy06} showed that tracing subhalos based on their mass at the time of accretion removes most of the antibias, and using this scheme, \citet{Conroy06} reproduced the clustering of the observed galaxies. Based on these results, we assume that satellites of LRG progenitors trace mass inside a halo; we cannot be more rigorous, as our mass resolution does not allow us to find subhalos or the mass of subhalos at any given time.

We have little information on how the progenitors of the LRGs are distributed at $z\gtrsim 1$, and therefore we test a wide range of initial $\mng$. We define 11 different HOD models at $z=1$ (\zoM) and at $z=2$ (\ztM) with variations in $\Ml/\Mm$ and $\al$, shown in Table \ref{tab:tHODz1} and \ref{tab:tHODz2}: at each redshift, 9 models with $\Ml/\Mm =$ 2, 10, and $\sim 25$, and $\al=$ 0.5, 1, and 2 and additional two models with only central galaxies. For Model 10 and Model 30, we assign central galaxies without the satellite term in Eq \ref{eq:Ngal}. For Model 11 and Model 31, we do not use Eq \ref{eq:Ngal} but consider a case in which halos in a very narrow range of mass can host central galaxies: we take a mass range of a factor 2.  We constrain the appropriate initial $\Ml$ and $\Mm$ by fitting the total number density of galaxies to $10^{-4}\itrihMpc$, the number density of the observed LRG sample with luminosity limit $-23.2 < M_g < -21.2$ at $0.16<z <0.36$ \citep{Zehavi05a}.  

We trace and locate the labeled particles (i.e., galaxies) in the dark matter halos at lower redshifts down to $z=0.3$ and derive their correlation function. We construct HODs of the evolved galaxies by counting the number of galaxies per halo at different halo mass bins.
We do not physically identify which of the galaxies become central or satellite galaxies in halos at low redshift, but we make a parametric distinction of the central and the satellite populations. That is, we estimate $\lan \ncen (M) \ran$ by counting halos with galaxies. When halos host more than one galaxy, we count the additional galaxies as satellites and calculate $\lan \nsat (M) \ran$. By construction, the central galaxies at low redshift follow the nearest integer distribution, while the probability distribution of satellites will be studied in the following sections. From the central and satellite number densities we derived, we fit the resulting $\mng$ to Eq \ref{eq:Ngal} and derive the best fit parameters: $\Mm$ from $\lan \ncen (M) \ran$ and $\Ml$ and $\al$ from $\lan \nsat (M) \ran$. The resulting $\mng$ is not necessarily in the exact form of Eq \ref{eq:Ngal}, but close enough in most cases that HOD parameters at the initial redshift and the final redshift can be compared. $\Ml$ and $\al$ are fitted over halo mass roughly larger than $\Ml$ to have a better description of the HOD for massive halos.

\subsection{Two-point correlation function}
The mapping between two-point correlation function and halo occupation distribution can be described analytically by a halo model of two components: a 1-halo term from the distribution of excess pairs within the same halo, which dominates the small-scale correlation function, and a 2-halo term from excess pairs between different halos, which dominates the large-scale correlation function \citep{Seljak00,Ma00,Berlind02,Cooray02}. The distribution of galaxy pairs within and between halos depends on the first and the second moments of HOD, given the halo mass function and the halo profile. Thus features in HOD are closely related to the 1-halo and 2-halo terms of the correlation function. 

From our simulations, the two-point correlation function of galaxies is calculated by counting the number of excess pairs at a given separation for each simulation box and then averaging spherically over all simulations. As we only study the distribution of pairs in this paper, we hereafter abbreviate `two-point correlation function' to `correlation function'.

\begin{figure*}[t]
\epsscale{1.9}
\figurenum{1}
\plotone{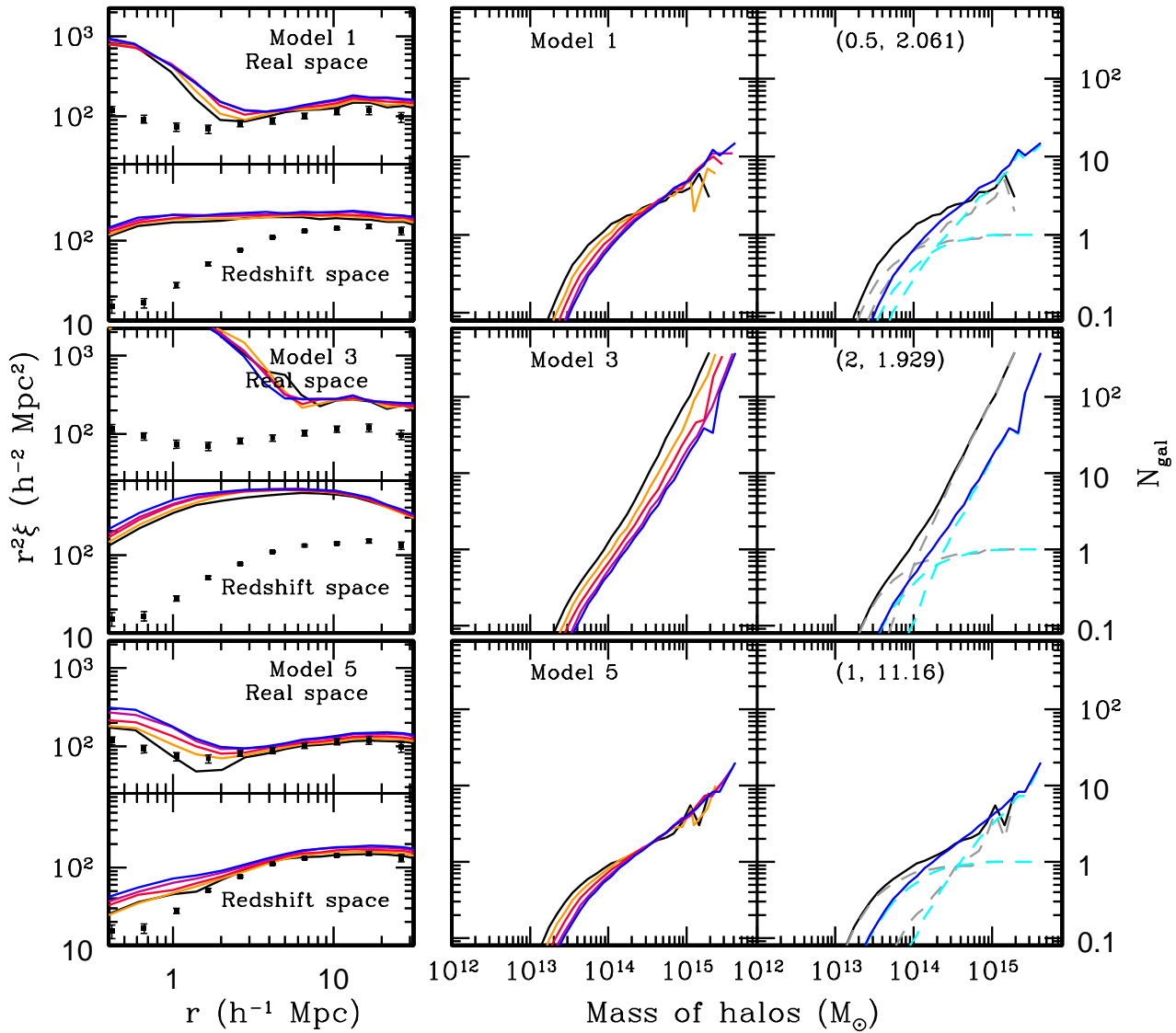}
\caption{The evolution of correlation functions and HODs of galaxies that flow passively from $z=1$. The left panels show $\xi$ in real space (upper) and $\xi$ in redshift space (lower). The data points are from \citet{Zehavi05a} for the observed clustering of the LRGs between $-23.2 < M_g < -21.2$ at $z=0.3$. The middle panels show the evolution of HODs, and the right panels show the decomposed HODs of central galaxies and satellite galaxies at $z=1$ (gray) and $z=0.3$ (cyan). Black line: the initial condition at $z=1$. Blue: at $z=0.3$. Red: at $z=0.8, 0.6, 0.4$. The input HOD for Model 11, which is a square function, appears smoothed due to our choice of mass bin. The values in the parenthesis in the right panels denote ($\al$, $\Ml/\Mm$) at $z=1$. }
\label{fig:fHODz1Mall}

\end{figure*}

\section{HOD and correlation function of passively flowing galaxies}\label{sec:z1Mz2M}
We start by summarizing the characteristics of passive flow evolution in the HOD, mainly the first moment $\mng$, and the correlation function. The clustering bias of these galaxies will converge toward unity with time on linear scales \citep{Dekel87,Nusser94,Fry96,Hamilton97,Tegmark98}, which means that the clustering of different models will converge with time\footnote{The expected bias at $z$, $b_{z}$, for passively flowing galaxies from $z_0$, is $b_{z}=(b_{z_0}-1)/G+1$ on linear scales where $b_{z_0}$ is the galaxy bias at $z_0$ and $G$ is the growth factor between $z$ and $z_0$. See \S~\ref{sec:BiasRed}.}. In this section, we will find how this translates to the way for halos to host galaxies and whether there is a similar convergence in correlation function on smaller scales. We assume galaxy populations that passively flow from $z=1$ and $z=2$. We use various HOD models that span from models with a significant satellite fraction to models with little or no satellite galaxies and study the difference in their fate. We also find how the result depends on the number density of galaxies (and hence the mass scale of the halos).

\begin{figure*}[t]
\figurenum{1}
\plotone{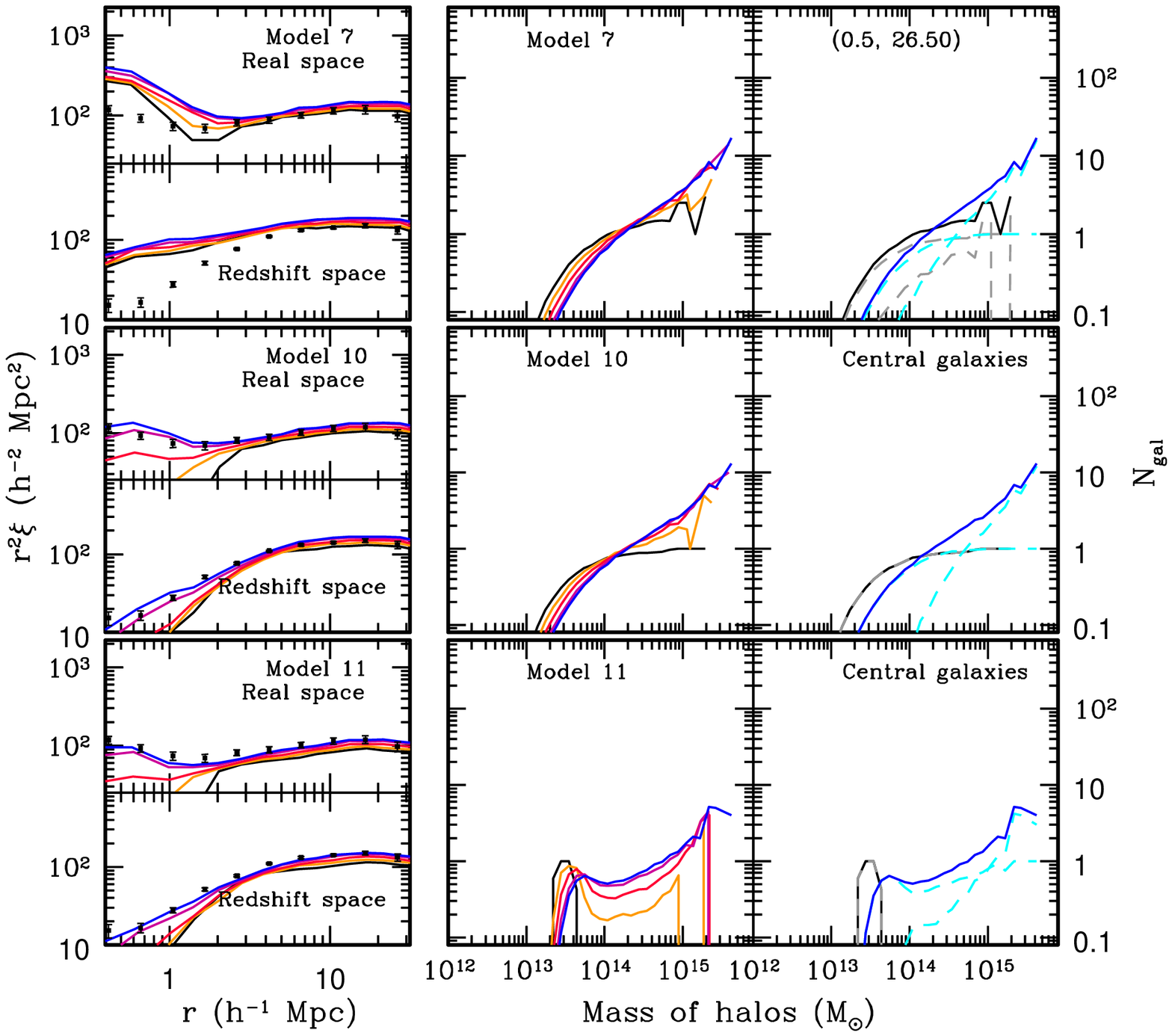}
\caption{continued}
\epsscale{1}
\end{figure*}

\begin{figure*}
\epsscale{1.6}
\figurenum{2}
\plotone{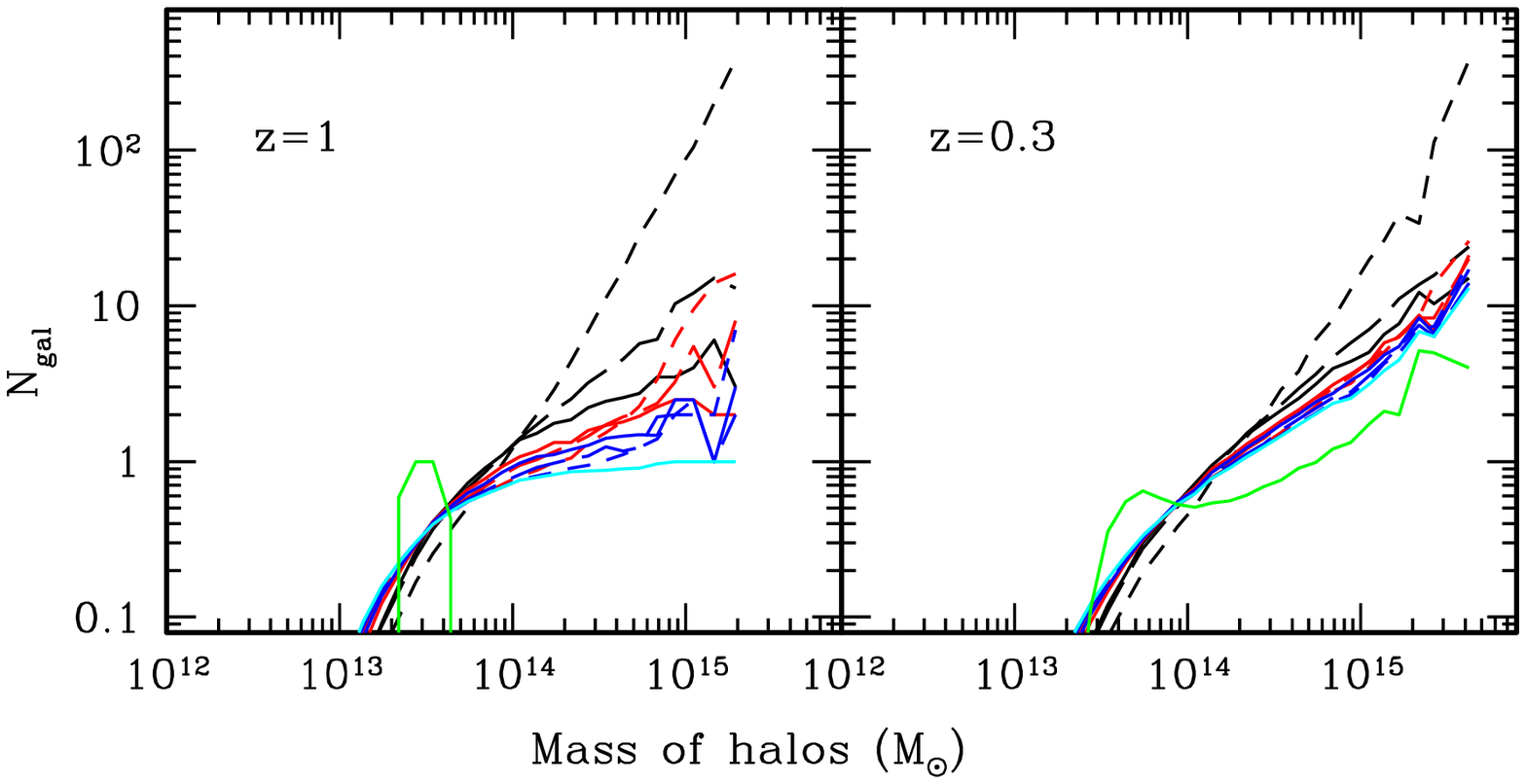}
\caption{The initial HODs at $z=1$ (left) and the final HODs at $z=0.3$ (right) for \zoM. Black : Models 1--3 (from Solid, long-dashed to short-dashed in order). Red : Models 4--6. Blue : Models 7--9. Cyan : Model 10. Green : Model 11. One finds that different initial $\mng$s asymptotically converge at $z=0.3$.  }
\label{fig:HODz1all}
\epsscale{1}
\end{figure*}

\begin{figure*}
\epsscale{1.6}
\figurenum{3}
\plotone{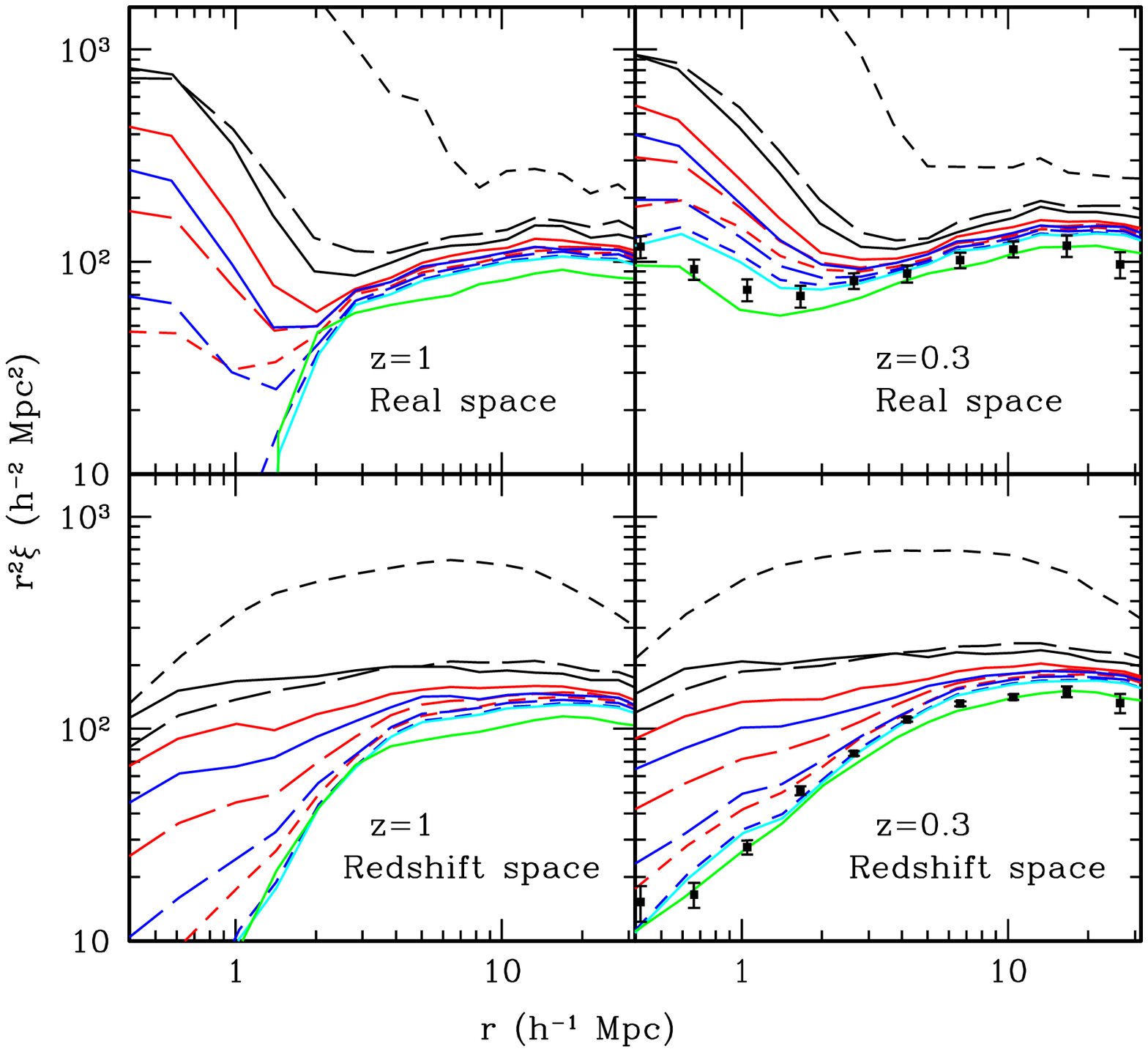}
\caption{The initial $\xi$ at $z=1$ (left) and the final $\xi$ at $z=0.3$ (right) for \zoM. Top : $\xi$ in real space. Bottom : $\xi$ in redshift space.  Black : Models 1--3 (from Solid, long-dashed to short-dashed in order). Red : Models 4--6. Blue : Models 7--9. Cyan : Model 10. Green : Model 11. }
\label{fig:Corz1all}
\epsscale{1}
\end{figure*}

\subsection{Galaxies flowing passively from $z=1$}\label{subsec:z1M}

We assign galaxies to dark matter halos at $z=1$ using 11 different HOD models (\zoM) and trace the resulting evolution of the HODs and correlation functions down to redshifts of 0.8, 0.6, 0.4, and 0.3. 

Table \ref{tab:tHODz1} lists the HOD parameters at $z=1$ and at $z=0.3$ for \zoM. Figure \ref{fig:fHODz1Mall} shows the evolution of correlation functions and HODs for Models 1, 3, 5, 7, 10, 11 among those listed in Table \ref{tab:tHODz1}. The black solid lines represent the initial HODs and correlation functions at $z=1$, and the blue solid lines are for the resulting HODs and correlation functions at $z=0.3$ while the reddish solid lines show the values at intermediate redshifts. The data points shown here are taken from \citet{Zehavi05a} for the observed clustering of the LRGs between $-23.2 < M_g < -21.2$ at $z=0.3$, which sets the fiducial number density of LRG progenitors in this paper.
As expected, the galaxy clustering (except for Model 3) grows slowly with time on all scales while the bias with respect to the dark matter decreases.

In Models 1, 2, and 3, in which $\Ml / \Mm \sim 2$, the fraction of satellite galaxies is much larger compared to those with larger values of $\Ml / \Mm$. As massive halos host more satellites, low-mass halos host a smaller number of central galaxies at a fixed number density. Thus the clustering is more weighted by high mass halos. The 1-halo term is prominent in these models because the large satellite fraction of massive halos emphasizes the structure of a halo. The trend becomes extreme in Model 3 in which massive halos are weighted most heavily. The evolution of clustering in Model 3 is not as evident as in other models, first because of the statistical noise due to the small number of very large halos and and second because of the large initial bias which drives a decrease in bias toward unity faster: as the galaxy clustering converges toward that of dark matter, the growth of galaxy clustering with a larger initial bias slows down more relative to the growth of dark matter clustering. As halos accrete mass and merge, the final $\mng$ at $z=0.3$ is parameterized with bigger values of $\Mm$ and $\Ml$ than at $z=1$ (Table \ref{tab:tHODz1}). The evolution of $\al$ shows an increase in Model 1, but changes very little in Models 2 and 3. In all cases, the fraction of satellites increases with time due to halo merging events.

Models 4, 5, and 6 start with $\Ml / \Mm \sim 10$ at $z=1$ with larger values of $\Ml$ and smaller values of $\Mm$, compared to Models 1--3. As a result, the clustering both from 1-halo and 2-halo terms is lower than in Models 1--3. While $\Mm$ increases with time, $\Ml$ changes little. The value of $\al$ increases or decreases toward near unity. Due to the steady $\Ml$, $\mng$ along different redshifts overlaps near $\Ml$. Model 5, with an initial $\al=1$, shows little evolution in $\al$.

In Models 7, 8, and 9, $\Ml / \Mm \sim 30$ initially. The clustering strength is even lower in these models as the values of $\Mm$ is even smaller. As the satellite population decreases from Models 7 to 9, the 1-halo term in the correlation functions is suppressed due to the exclusion effect in a halo finder. However, the evolution of clustering quickly recovers the exclusion effect at later redshifts. The value of $\Ml$ decreases with time while $\Mm$ increases with time. Again, $\al$  evolves toward near unity. 

In Model 10, we initially populate only the halos with central galaxies without satellites. At low redshift, many of these central galaxies turn into satellite galaxies by halo merging, and the results at $z=0.3$ are very similar to those of Model 9. 

The evolution of HOD for Models 1--10 is therefore characterized as following: $\al$ approaches toward an attractor near unity, and a large initial $\Ml/\Mm$ decreases with time (due to halo merging without galaxy merging) while a small initial $\Ml/\Mm$ stalls. Such evolution of HOD parameters drives the convergence of $\mng$ at low redshift. Figure \ref{fig:HODz1all} and \ref{fig:Corz1all} show the comparisons of HOD and correlation functions between different models at the initial redshift ($z=1$) and the final redshift ($z=0.3$). From the figures, it is clear that passive flow evolution leads to an asymptotic convergence of clustering on scales above $r \sim 3 \hMpc$ and of the first moment of HOD. Most of the models produce a stronger clustering than that of the observed LRG data while Model 10 marginally fits the data. From Figure \ref{fig:Corz1all}, clustering on small scales ($r \lesssim 3 \hMpc$) varies considerably despite the similarity in their HODs. The difference on small scales implies that the second moment of HOD, that is, the average number of galaxy pairs within a halo, may differ for different models. The difference in redshift-space clustering at $z=0.3$ extends to the larger scale than in real-space clustering due to the finger-of-God effect. We revisit this issue of second moments in \S~\ref{sec:z1Mz2MNN}.

\begin{figure*}[t]
\epsscale{1.9}
\figurenum{4}
\plotone{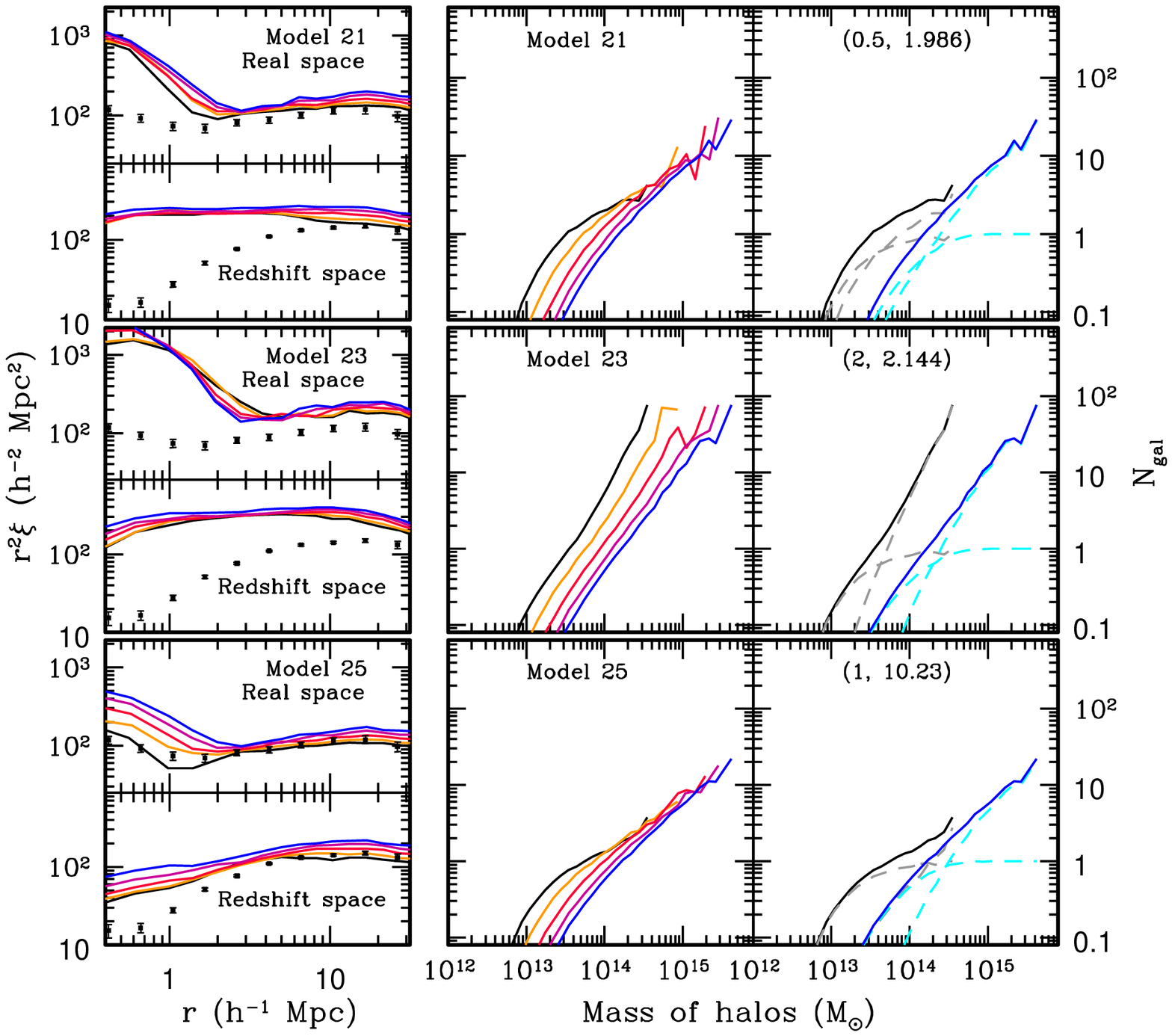}
\caption{The evolution of correlation functions and HODs of galaxies that flow passively from $z=2$. The left panels show $\xi$ in real space (upper) and $\xi$ in redshift space (lower). The data points are from \citet{Zehavi05a} for the observed clustering of the LRGs between $-23.2 < M_g < -21.2$ at $z=0.3$. The middle panels show the evolution of HODs, and the right panels show the decomposed HODs of central galaxies and satellite galaxies at $z=2$ (gray) and $z=0.3$ (cyan). Black line: the initial condition at $z=2$. Blue: at $z=0.3$. Red: at $z=1.5, 1, 0.6$. The input HOD for Model 31, which is a square function, appears smoothed due to our choice of mass bin. The values in the parenthesis in the right panels denote ($\al$, $\Ml/\Mm$) at $z=2$.}
\label{fig:fHODz2Mall}
\end{figure*}

\begin{figure*}[t]
\figurenum{4}
\plotone{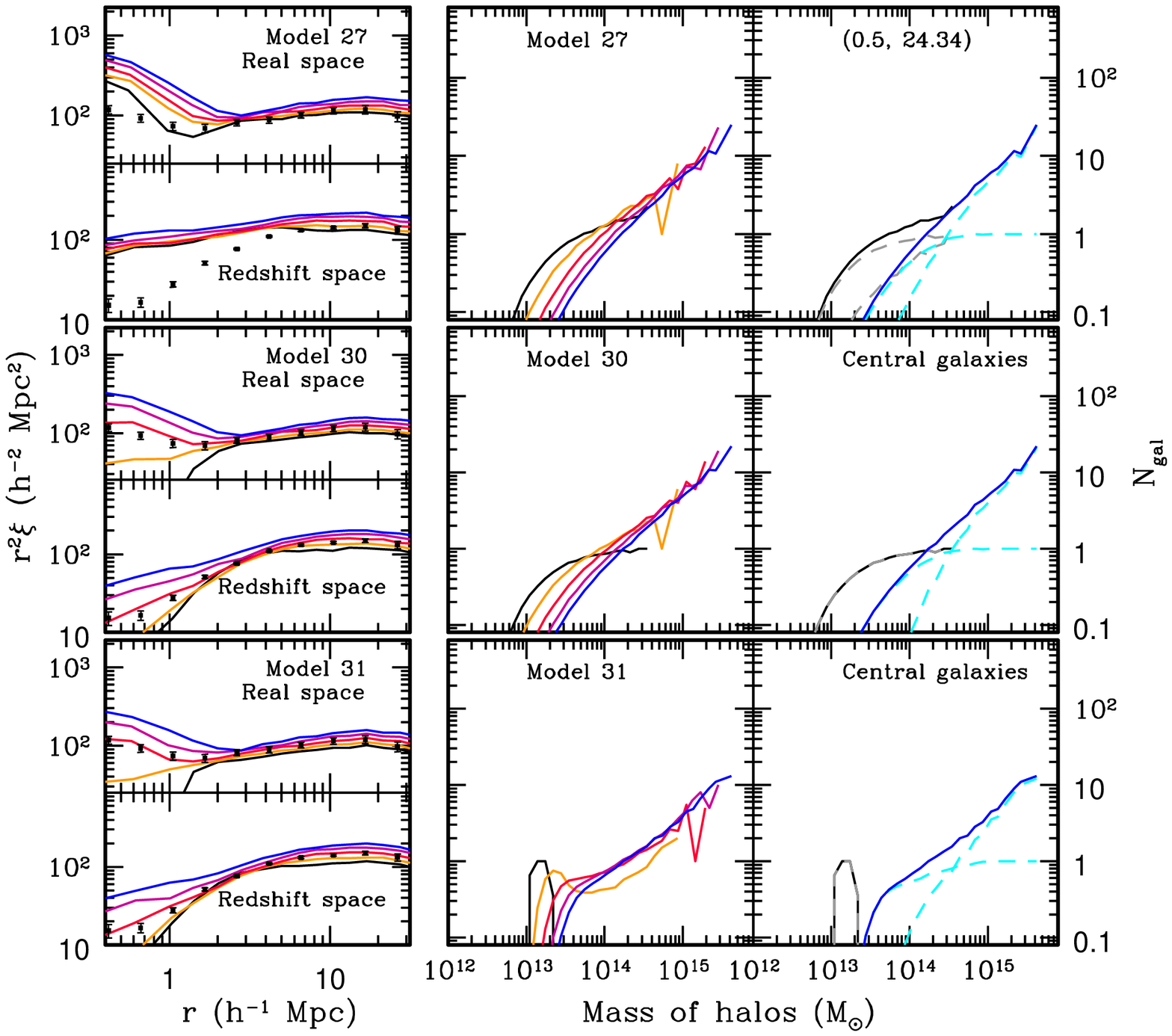}
\caption{continued}
\epsscale{1}
\end{figure*}

In all models, the increase in $\Mm$ is consistent with the typical mass accretion between $z=1$ and $z=0.3$ for the corresponding mass range \citep[e.g.,][]{Wech02}. On the other hand, the mass accretion of halos near $\Ml$ will often involve accretions of new galaxies into the halos. For Models 1--3, the galaxy accretion near $\Ml$ is minor compared to the mass accretion, as the initial value of $\Ml$ at $z=1$ is closer to $\Mm$, driving the increase in $\Ml$ with time. For Models 7--9, the large $\Ml/\Mm$ at $z=1$ implies that the mass accretion to a halo near $\Ml$ will often accompany galaxy accretion, which must be efficient enough to decrease $\Ml$ with time. Models 4--6 shows an intermediate behavior. 
\\

The small values of $\Ml/\Mm$ at low redshift, that is, the lack of shoulder between $\Ml$ and $\Mm$, is different from the predictions of $\mng$ for subhalos or galaxies in non-passive studies \citep[e.g.,][]{Berlind03, Krav04, Conroy06, Weinberg06}). The discrepancy is a natural result for passive flow evolution as we do not include any merging or tidal disruption of tracers. With tidal stripping and merging between subhalos or merging to the center, the number of subhalos will decrease with time for a broad range of host halo mass. Semi-analytic studies and N-body simulations show that the resulting cumulative number density of subhalos reaches unity at a small subhalo-to-host halo mass ratio (or circular velocity ratio) \citep[e.g.,][]{Zentner03,Taylor04,Oguri04,Diemand04,Gao04b,vanB05b,Zentner05}. This small ratio should roughly represent a large value of $\Ml/\Mm$ by the definition of $\Mm$ and $\Ml$. Conversely, the cumulative subhalo mass function with no tidal disruption and merging \citep[e.g.,][]{Zentner03,vanB05b} hints small values of $\Ml/\Mm$ for our passively flowing case. Thus, one can reason that, with non-passive flow, $\Ml$ would increase, and then, for fixed number density, $\Mm$ would have to decrease to include new central galaxies (i.e., host halos that have accreted enough mass to pass the threshold). With decreasing $\Mm$ (i.e., decreasing mass threshold for subhalos), $\Ml$ will be readjusted. However the value of $\Ml/\Mm$ will remain large, as the subhalo mass function is roughly self-similar for different host halo mass, although slightly more abundant in more massive host halos \citep[e.g.,][]{Gao04b,vanB05b,Zentner05}.

For galaxies, tidal disruption will be much less efficient than for subhalos as they are more tightly bound systems. Nevertheless, dynamical friction will eventually merge some of them to the central object, more effectively for satellites sitting in more massive subhalos \citep{Binney87}, which will result in a larger $\Ml/\Mm$ than our passively flowing case. The merging products corresponding to the discrepancy between passively and non-passively flowing galaxies should have consequences elsewhere, such as in LRG luminosity function, so that we can verify the model.

As a caveat, we note that our passively flowing samples at the observed time might not simply correspond to galaxy populations in non-passive studies, which are constructed by relating the luminosity of galaxies to the subhalos and host halos above a given mass (or velocity) cut, whether the selection of the subhalos is based on values at the time of accretion or at the observed time. Again, our samples are defined at the initial time, not at the observed time.

All the models we discussed so far show the composite information on galaxy statistics from halos of various mass. To probe the response of the delta function in host-halo mass to passive flow evolution, we select a mass range with a width $2$ which gives the fiducial number density and assign one central galaxy per each halo (Model 11). This scheme populates more galaxies in lower mass halos than the previous models do, depriving massive halos of galaxies at the initial redshift. Due to the deprivation of galaxy-hosting halos of mass between $\Mm$ and $\Ml$, relative to the other models, this model mimics the observed shoulder in HOD of the LRGs at $z=0.3$. The resulting correlation function at $z=0.3$ appears consistent with the observed data for LRGs as well. The response of a delta function at different mass will slightly vary; otherwise, a superposition of delta functions of different halo-mass, when weighted by the initial $\mng$, will correspond to the evolution of galaxies in Model 10. The correlation function of Model 11 does not appear different from other models as much as its HOD does.

\subsection{Galaxies flowing passively from $z=2$}
\begin{figure*}
\epsscale{1.6}
\figurenum{5}
\plotone{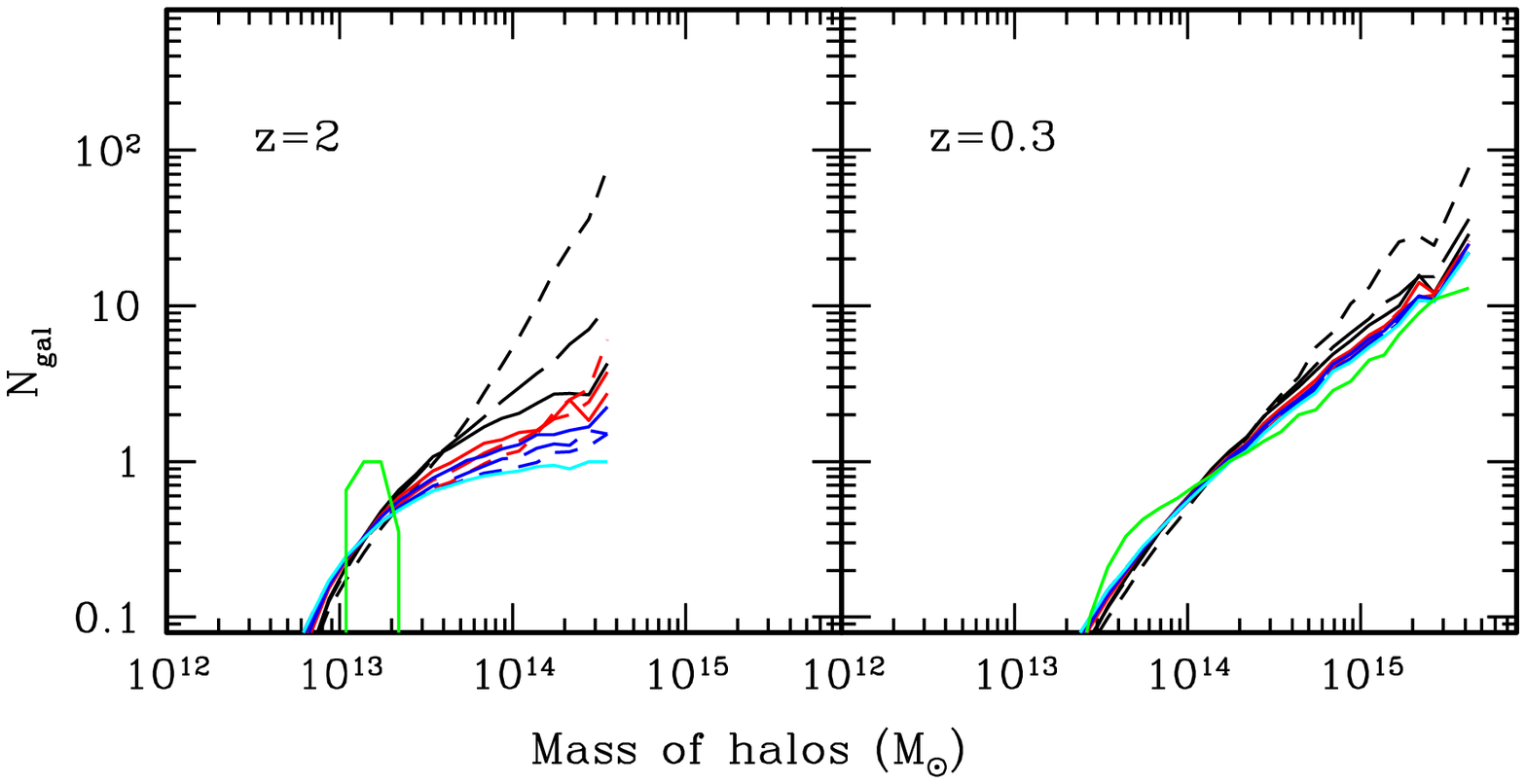}
\caption{The initial HODs at $z=2$ (left) and the final HODs at $z=0.3$ (right) for \ztM. Black : Models 21--23 (from Solid, long-dashed to short-dashed in order). Red : Models 24--26. Blue : Models 27--29. Cyan : Model 30. Green : Model 31. One finds that the asymptotic convergence of $\mng$ is stronger when galaxies flow passively from $z=2$ than from $z=1$.  }
\label{fig:HODz2all}
\end{figure*}

\begin{figure*}
\figurenum{6}
\plotone{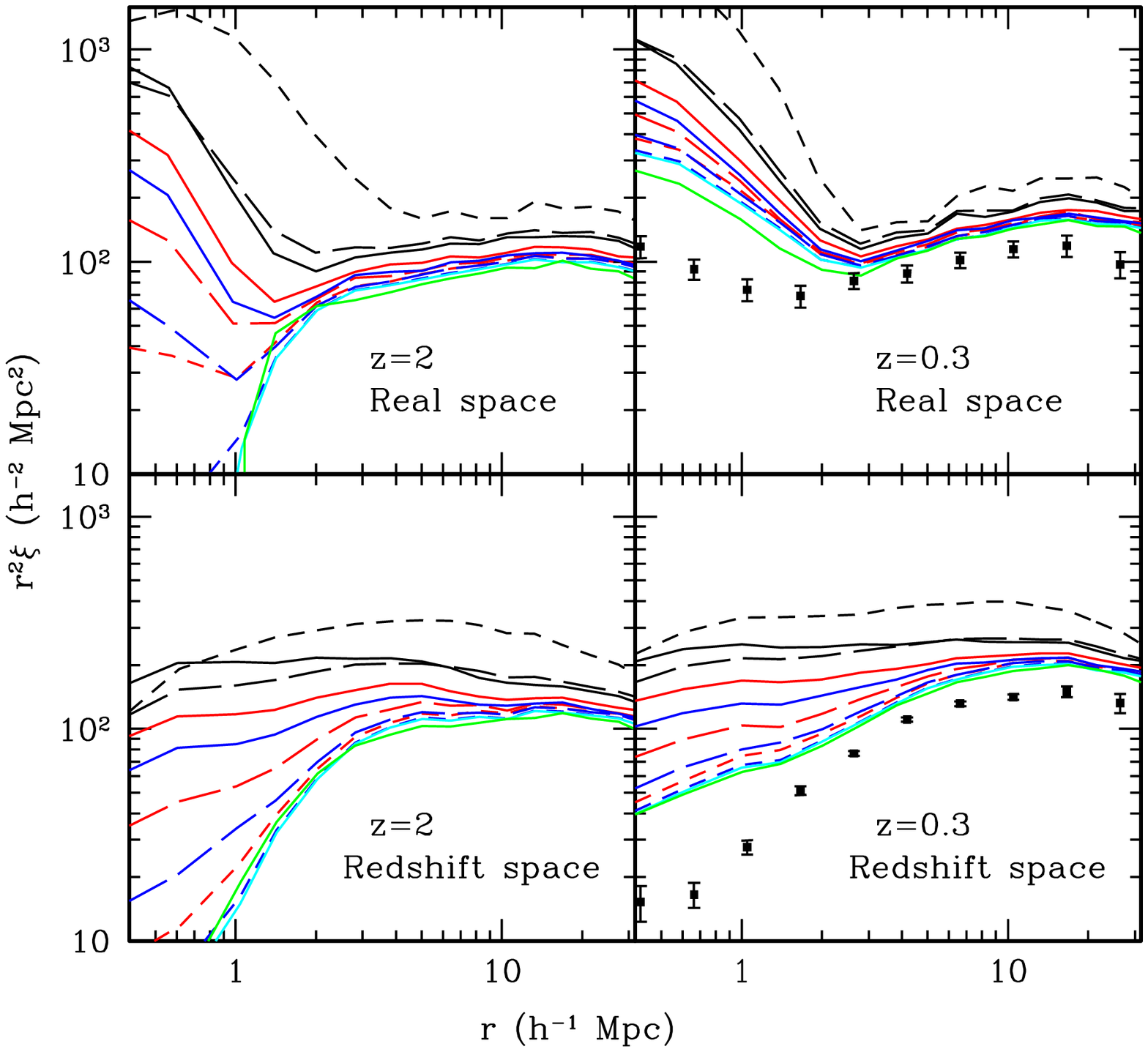}
\caption{The initial $\xi$ at $z=2$ (left) and the final $\xi$ at $z=0.3$ (right) for \ztM. Top : $\xi$ in real space. Bottom : $\xi$ in redshift space.  Black : Models 21--23 (from Solid, long-dashed to short-dashed in order). Red : Models 24--26. Blue : Models 27--29. Cyan : Model 30. Green : Model 31. The convergence in correlation function is stronger when galaxies flow passively from $z=2$ than from $z=1$. }
\label{fig:Corz2all}
\epsscale{1}
\end{figure*}

We next observe the response of correlation function to passive flow from higher redshift. We populate galaxies at $z=2$ for \ztM\ and compare the resulting evolution of correlation function and HODs at $z=0.3$ (Figure \ref{fig:fHODz2Mall} and Table \ref{tab:tHODz2}) with those from $z=1$ (\zoM). We investigate whether and where the convergence of HOD parameters occurs if galaxies flow passively from $z=2$, relative to the galaxies from $z=1$. We will show that the evolution from $z=2$ to $z=0.3$ is qualitatively similar to the evolution from $z=1$ to $z=0.3$ except that the consecutive time steps cover a broader range of evolution. 

At higher redshift, the typical halos have a lower mass. The average mass of host halos for the fiducial number density is smaller by more than a factor of 2 than that at $z=1$, but these halos are more biased than the typical host halos at $z=1$ due to the rarity of the halos at high redshift \citep{PS74,BBKS86}. The initial clustering strengths at $z=2$ are similar to those from $z=1$, and so the resulting correlation function at $z=0.3$ is larger for \ztM\ than \zoM.  

In general, the transition from the 1-halo to the 2-halo term in the correlation function occurs at a smaller separation at the initial redshift, as expected from the smaller halo-size of the dominant halo population and the steeper halo mass function at $z=2$.  \ztM\ start with a similar or slightly smaller initial ratio of satellite to central galaxies compared to \zoM\ except for Model 23 and Model 3. However, \ztM\ produce a larger satellite fraction at $z=0.3$, which is due to the longer time available for halo merging and accretion events.

In Model 21, 22, and 23, in which $\Ml / \Mm \sim 2$, the fraction of satellite galaxies is much larger compared to the other models, like in Models 1-3. Both $\Mm$ and $\Ml$ increase at $z=0.3$, and $\al$ approaches somewhere near unity.

Model 24, 25, and 26 have initial $\Ml / \Mm \sim 10$. Both $\Mm$ and $\Ml$ increase with time, unlike Models 4-6, but the ratio of $\Ml/\Mm$ decreases considerably at $z=0.3$, as in Models 4-6. Again, the value of $\al$ converges toward near unity. 

In Model 27, 28, and 29 with $\Ml / \Mm \sim 25$ initially, we observe the strong exclusion effect at $z=2$ that quickly recovers with time. In these models, $\Ml$ at $z=0.3$ is less than $\Ml$ at $z=2$ just like in the corresponding models at $z=1$ (Models 7--9). Again, $\al$ approaches toward unity.

Model 30 is almost identical to Model 29. Model 31 evolves to have a region of shoulder at $z=1$ but then takes on a power-law shape at $z=0.3$ and produces a stronger clustering than the observed LRGs.

\begin{figure*}[t]
\epsscale{1.6}
\figurenum{7}
\plotone{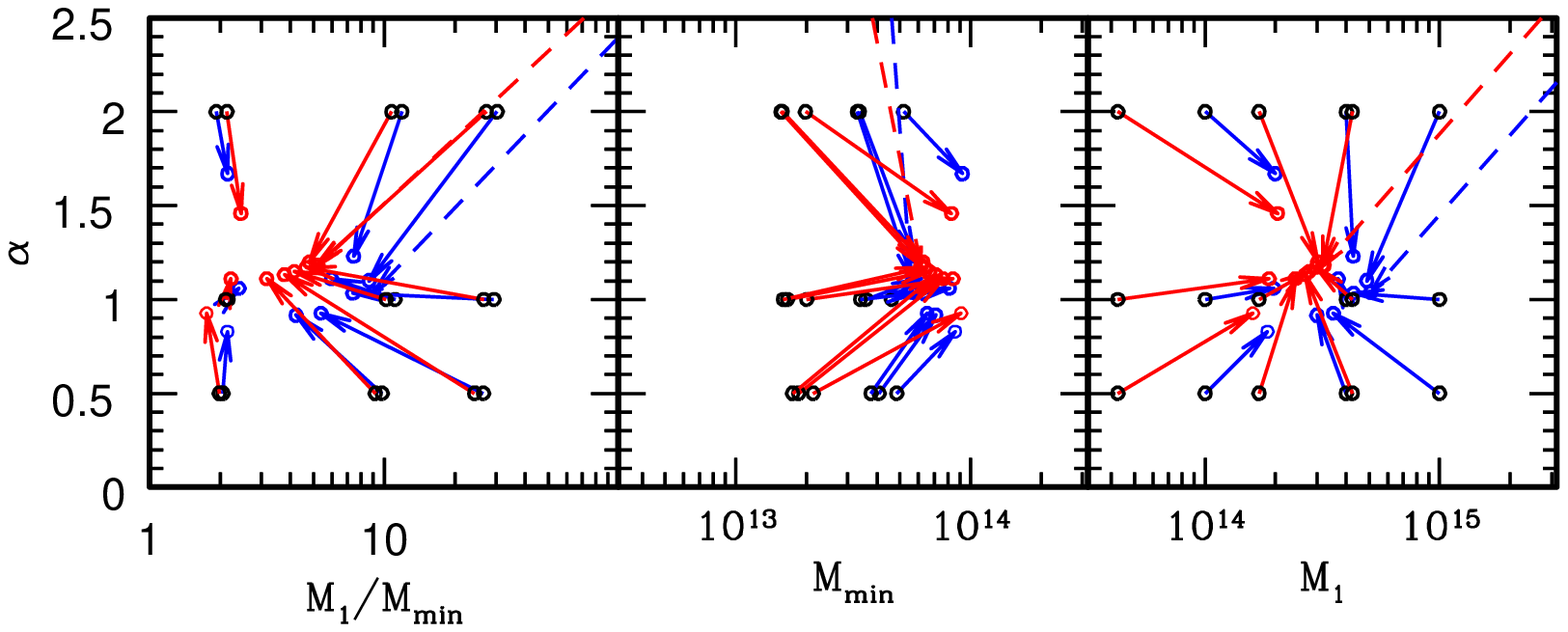}
\caption{The evolution of passively flowing galaxies in the HOD parameters. Arrows denote the direction of evolution. Black points : initial HODs of Models 1--10 (at $z=1$) and Models 21--30 (at $z=2$). Blue points : the final HODs at $z=0.3$ for Models 1--10. Red points : the final HODs at $z=0.3$ for Models 21--30. The dashed lines are for Model 10 and 30 for which we cannot define the initial $\Ml$ and $\al$. One finds that for passively flowing populations, a large $\Ml/\Mm$ in general decreases with time and $\al$ approaches toward near unity.}
\label{fig:Phase}
\epsscale{1}
\end{figure*}

From the evolution of the real-space correlation, we find a feature prevailing in most of the models: the growth of clustering is impeded near $r\sim 2-5 \hMpc$ (but $r>3 \hMpc$ for Model 23). That is, in Figure \ref{fig:fHODz2Mall}, the correlation functions at different redshift become squeezed together over $r\sim 2-5 \hMpc$, implying a suppression in growth of galaxy clustering. This could be a signature associated with a turn-around and infall in the structure formation. In redshift space, the features are less obvious but still traceable. For \zoM, the corresponding feature is observable on slightly larger scales. We further discuss about this feature in \S~\ref{sec:evolCor}.

Figure \ref{fig:HODz2all} and \ref{fig:Corz2all} show the comparisons of HOD and correlation function for \ztM\ at the initial redshift ($z=2$) and the final redshift ($z=0.3$). Figure \ref{fig:HODz2all} shows a stronger convergence of these final HODs than observed in \zoM. The asymptotic $\al$ at $z=0.3$ moves toward slightly larger $\al$ than in \zoM (Figure \ref{fig:HODz2all}). Likewise, Figure \ref{fig:Corz2all} shows a stronger convergence between models in the large-scale clustering ($r \gtrsim 3\hMpc$) at $z=0.3$ as well as in small-scale clustering, which is due to the larger bias at the initial redshift and longer evolution time. Again, the small-scale correlation function remains different between models although their HODs look nearly identical.

\begin{figure*}[t]
\epsscale{1.6}
\figurenum{8}
\plotone{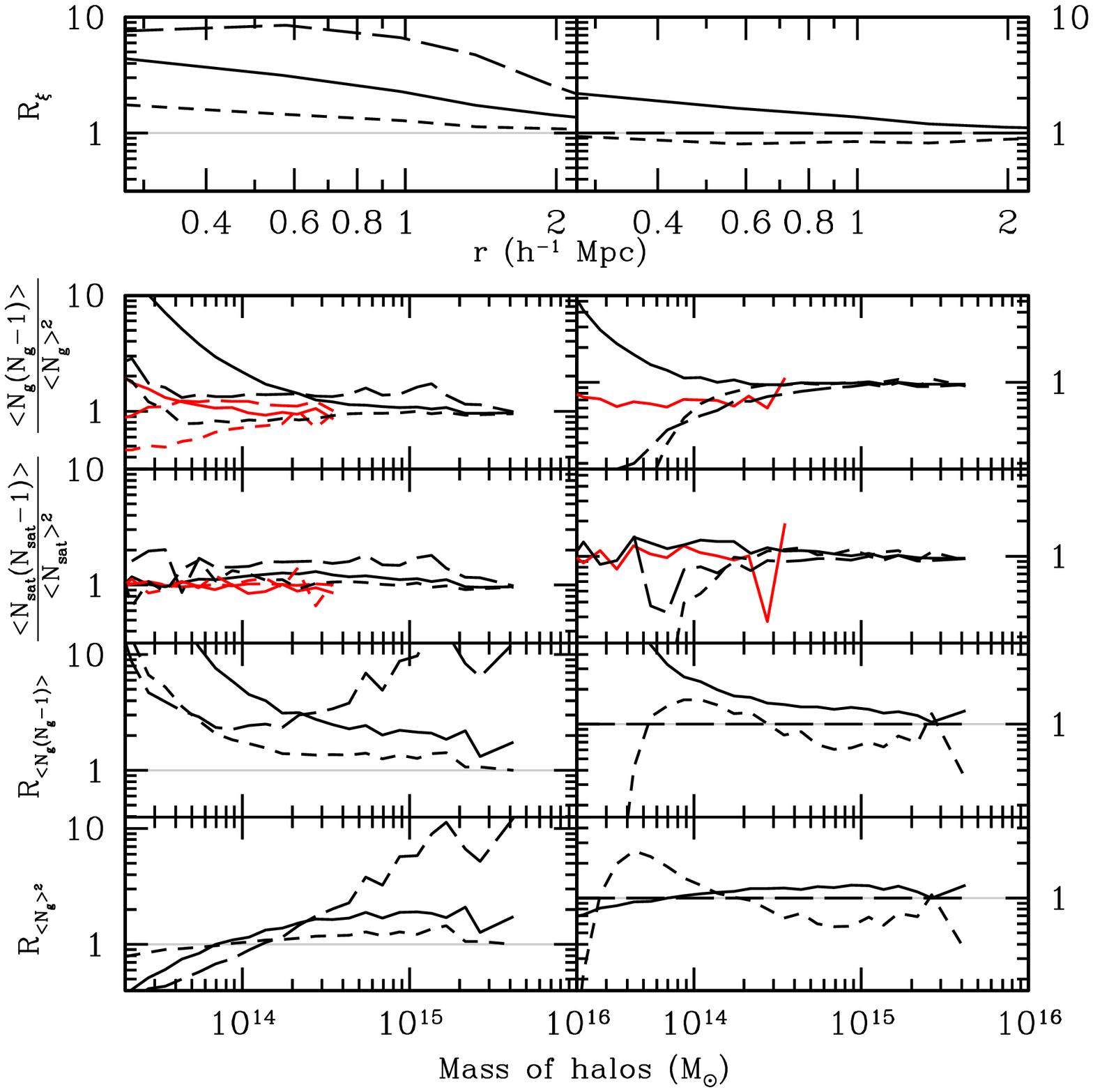}
\caption{The comparisons of $\NNs$ and real-space $\xi$ of galaxies at $z=0.3$ for Models 21, 23, and 25 (left) and 27, 30, and 31 (right). Three different models in each panel are distinguished by different line types : solid lines for Models 21 and 27, long-dashed for Models 23 and 30, and short-dashed for Models 25 and 31. $R_\xi$, $R_{\NNs}$, and $R_{\mng^2}$ are the ratios of the model quantities to those from Model 30 (for example, $R_\xi = \xi_{\rm Model\;X}/\xi_{\rm Model\;30}$). In the second and the third fields, red lines are for $z=2$, and black lines are for $z=0.3$. In the third field, $\NNss$ and $\mngs$ for satellites are calculated among halos with central galaxies. We find that the galaxy populations asymptotically converge to Poisson at low redshift for massive halos. The difference in small-scale clustering appears consistent with the difference in the average pair counts within a halo.}
\label{fig:NNsz2all}
\epsscale{1}
\end{figure*}

Figure \ref{fig:Phase} provides a clear view of similarities and differences in the evolution of HOD parameters for the two formation redshifts. The middle panel shows that $\Mm$ increases with time while $\al$ converges toward near unity. The range of resulting $\Mm$ at $z=0.3$ for \ztM\ is similar to the range of $\Mm$ at $z=0.3$ for \zoM. The right panel shows that $\Ml$ tends to converge to $\Ml \sim 2-5 \times 10^{14} \Msun$ at low redshift while the convergence is stronger and at lower $\Ml$ for \ztM. The left panel shows that the resulting $\Ml/\Mm$ converges toward $\Ml/\Mm \sim 3-4 $ while the convergence is stronger and at lower value of $\Ml/\Mm$ for \ztM. It appears that the convergence of $\al$ is stronger and at a slightly larger value, somewhere above unity, for \ztM\ than \zoM.

\subsection{A test with host halos of lower mass: increasing the number density}\label{subsec:z1n4}
We have shown the characteristics of passively flowing galaxies for high mass halos that are consistent with the number density of the LRGs between $-23.2 < M_g < -21.2$ at $z=0.3$. The characteristics could depend on the peculiarity (or rarity) of an extreme tail of nonlinearity, and thus we investigate the effect of passive flow for the galaxies of a larger number density, i.e., in lower mass halos. As both \zoM\ and \ztM\ produced a clustering stronger than one observed for the LRGs, we also ask whether the observed LRG clustering can be better explained as a random fraction of a parent population with a larger number density that has evolved passively to lower redshift. 

We assign galaxies to dark matter halos at $z=1$ using values of $\Ml/\Mm$ and $\al$ similar to Models 7--10 but with {\it four} times the fiducial number density ($=4\times 10^{-4}\itrihMpc$) and trace the resulting evolution of $\mng$ and two-point correlation functions down to redshifts of 0.8, 0.6, 0.4, and 0.3 (Table \ref{tab:tHODz1n4}). We label these models as \zoMfn. These models have slightly larger initial and final satellite fractions compared to Models 7--10.

The evolved clustering of \zoMfn\ is weaker at $z=0.3$ than the clustering of the LRGs, which implies that the parent population of the observed LRGs will have a number density between $10^{-4}\itrihMpc$ and $4\times 10^{-4}\itrihMpc$ under passive flow evolution. The evolution of $\Ml/\Mm$ and $\al$ for these models is very similar to that for \zoM, although the convergence of $\Mm$ and $\Ml$ is at lower mass than in \zoM. Therefore the characteristics of passive flow evolution in HOD parameters we find, that is, the decrease in $\Ml/\Mm$ and $\al$, are fairly robust for a wide range of halo mass. 
\\

The convergence of $\Ml/\Mm$ and $\al$ can be understood from the idea that the passive flow drives the galaxies to flow and distribute like mass, as gravity does not distinguish galaxies from other mass components. The distribution of passively flowing galaxies will then converge to $\lan N({\rm M})\ran$ of mass, which is a power law with an index of unity. With our parameterization, the convergence of $\mng$ to the linear equation in halo mass forces a decrease in $\Ml/\Mm$ and convergence of $\al$ toward slightly larger than unity due to our division to central and satellite populations. This asymptotic convergence will be less efficient in mass components lower than $\Mm$, as host halos {\it build up} hierarchically to larger ones with time rather than disassemble in field, therefore maintaining the biased clustering. The normalization of $\mng$ will be determined by distributing the total number of galaxies to halos of $M \gtrsim \Mm$.  
 
Note that HODs of all models except for Model 11 resemble the old galaxy population defined at the observed redshifts (i.e., without passive-flow restriction) in studies of SPH simulations or semi-analytic models \citep{Berlind03,Zheng05}. Model 11 on the other hand resembles the young galaxy populations in those studies.

\section{The evolution of pair counts within a halo}\label{sec:z1Mz2MNN}

In the previous sections, we found that while various initial models evolve to a well-defined region of $\mng$ at $z=0.3$, there still remains a considerable difference in the small-scale clustering at $z=0.3$ between models. While the clustering on large scales can be modeled by the 2-halo term that depends on the excess number of pairs between halos, i.e., $\mng$ weighted by halo bias and halo mass function, the small-scale clustering (1-halo term) depends on the average counts of excess pairs within a halo, i.e., $\NNs$, the spatial distribution of the pairs within the halo, and halo mass function. Therefore the comparison of pair counts within a halo will help to explain the difference in the small-scale correlation functions. 

When samples are defined at the observed redshifts, SPH and semi-analytics calculations imply that the satellite probability distribution is modeled well by the Poisson distribution \citep{White01,Berlind02,Krav04,Zheng05} that gives $\NNs = \mng^2$, while the central galaxies follow $\NNs = 0$. From the pair counts of the passively flowing galaxies, we thus will find the effect of removing processes, such as merging, destruction, or creation of galaxies, on the second moments of HOD, again, provided that the initial populations follow HOD as a function of halo mass only and have Poisson-distributed initial satellites if any.

Figure \ref{fig:NNsz2all} shows the pair counts at $z=0.3$ for galaxies that passively flowing from $z=2$ (Models 21, 23, 25, 27, 30, and 31) and the corresponding small-scale correlation function in real space. In the second field of the panels, the pair counts are divided by $\mng^2$ to be compared to the nearest integer and the Poisson distribution: $\NNsN=0$ for the nearest integer distribution and unity for the Poisson distribution. For HOD models with a negligible satellite fraction for halos at the low-mass end where $\lan \ncen \ran < 1$, $\NNsN$ will start from zero and approach unity as the satellite fraction grows with $M$ (such as Models 26, 28, and 29). On the other hand, some of our models allow a non-zero satellite fraction for $\lan \ncen \ran < 1$ at the initial redshifts, and this reverses the shape of $\NNsN$ at the low-mass end as shown in Figure \ref{fig:NNsz2all} for Models 1 and 7.

\begin{figure*}[t]
\epsscale{1.6}
\figurenum{9}
\plotone{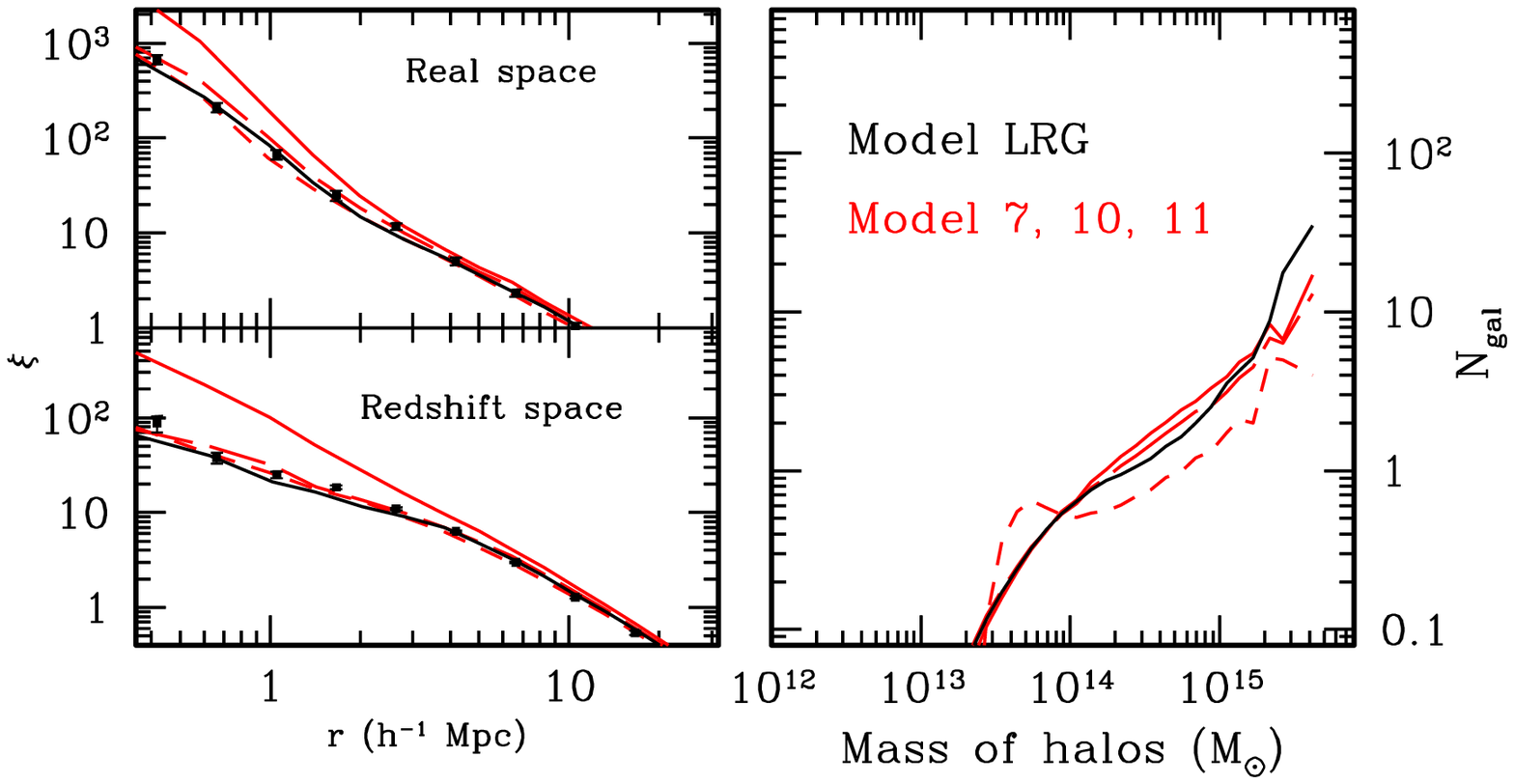}
\caption{An example of our HOD fits to the observational LRGs. The resulting $\mng$ and correlation function (black lines) is compared with the observed correlation function for LRGs from \citet{Zehavi05a} (black points) and Models 7, 10, 11 (red lines). Left : correlation functions in real space (top) and in redshift space (bottom). Right : $\mng$ at $z=0.3$. Model 7 : solid lines. Model 10 : long-dashed lines. Model 11: short-dashed lines.  }
\label{fig:Corallz0.3M}
\end{figure*}

From the figure, the evolved $\NNsN$ at $z=0.3$ nicely converges to unity for massive halos in all models. This implies that the passive flow evolution results in satellite populations that follow a Poisson distribution over a wide range of halo mass, although some of the models, such as Models 29, 30, and 31, started with a nearly or purely nearest integer distribution at $z=2$. In low-mass halos, galaxies at $z=0.3$ follow a nearest integer distribution by construction (\S~\ref{subsec:models}) as long as each halo hosts one or zero central galaxy. The third field of the panels, where we show the pair counts of satellites alone calculated among halos with central galaxies, reveals further details on satellite distributions. Except for Model 31, satellites in models with a smaller initial satellite fraction or a larger increase in the satellite fraction between $z=2$ and $z=0.3$ (i.e., Model 26, 28, 29, 30), converge better to the Poisson distribution, though slightly sub-Poisson. Satellites in models with a larger initial satellite fraction and those in Model 31 tend to be slightly super-Poisson, while converging toward a Poisson distribution at the massive end.

Considering the Poisson distribution of subhalos or satellite galaxies from non-passive evolution \citep{White01,Berlind02,Krav04,Zheng05}, our results implies that the processes of merging, formation, or destruction of galaxies (or subhalos), which are not included in our passive flow evolution schemes, preserve the Poisson distribution. This implies that the convergence toward the Poisson distribution probably arises from the random nature of the halo merging (Zheng Zheng, a private communication). Although not shown in the figure, Models 1--10 show similar results while the convergence toward the Poisson statistics is not as strong as \ztM. In Model 11, all galaxies for $M>10^{14} \Msun$ approach a Poisson distribution (i.e., $\NNsN \sim 1$), while being slightly super-Poisson, due to the relatively large satellite-to-central galaxy ratio for intermediate mass halos despite the small total satellite fraction. Model 31 has a more moderate transition from central to satellite domination.

The fourth and the fifth fields in Figure \ref{fig:NNsz2all} show ratios of $\NNs$ and $\mng^2$ with respect to the values for Model 30. When we compare the first and the fourth fields of the figure, we find that the trend of $\NNs$ for $M\gtrsim 10^{14}\Msun$ is qualitatively consistent with the trend of $\xi$ on small scales, even without considering the details of the mass-dependent halo profile. The comparison between $\NNs$ and $\mng^2$ shows that the difference in $\NNs$ is rooted in the small difference in $\mng^2$ between models (see \S~\ref{sec:z0.3H}). The overall convergence of $\mng$ gives the convergence in clustering, and it is stronger on large scales because the 2-halo term depends on the integrated effect of $\mng$ over halo mass. The small-scale clustering is more sensitive to the details of $\mng$ through $\NNs$, as each mass bin corresponds to a range of scale $r$ in $\xi(r)$ consistent with a typical halo size of that mass. We conclude that the difference in small-scale clustering between models is due to the difference in $\mng$ (and its effect on $\NNs$), and not to the nature of the scatter (e.g., Poisson or non-Poisson).

\section{Comparisons to the current observation of the LRGs}\label{sec:z0.3M}
Previous sections show that passive flow leads to a small value of $\Ml/\Mm$ and $\al$: in general, $\Ml/\Mm < 10$ and $\al$ slightly above unity. In this section, we compare our HOD parameters for the passive flow with the fit to the observed LRGs by \citet{Zheng07}. Even for the same cosmology, the details of the group finding method or the resolution of simulation may alter mass scale or mass function of halos. Therefore, in order to discuss the differences between our results and the fit by \citet{Zheng07}, we first need to identify an LRG population at $z=0.3$ in our own simulations that is consistent with the observed clustering and the shape of $\mng$ adopted by \citet{Zheng07}. By this, we can  calibrate our mass scale and assess the physical significance of the parameter space of the HOD confined by passive flow evolution.

A five-parameter HOD fit (eq. [\ref{eq:hod5param}]) to the observed LRG clustering, derived for our fiducial cosmology, is kindly provided by Zheng Zheng (in the footnote of Table \ref{tab:tHODz0.3}). The five-parameter HOD is defined as
\begin{eqnarray}
\mng &=& 0.5 \left[ 1+\erf [\log_{10}(M/\Mm')/\sigma_M] \right] \nonumber \\
&\times& \left[ 1+[(M-M_0)/\Ml']^\al \right]
\label{eq:hod5param}
\end{eqnarray}
where $\Mm'$ is the characteristic minimum mass to host a central galaxy, $\Ml'$ is a mass for a halo with a central galaxy to host one satellite when $M_0 \ll \Ml'$, $M_0$ is the truncation mass for satellites, and $\sigma_M$ is the characteristic transition width  \citep{Zheng05}.

Compared to the Jenkins mass function \citep{Jenkins01} used by \citet{Zheng07}, our halo mass function produces slightly more halos for $M < 2\times 10^{15}\Msun $, which can be corrected by rescaling our halo mass down by $\sim 5\%$, and less halos for $  2\times 10^{15}\Msun <M < 3\times 10^{15}\Msun$, which is possibly due to Poisson noise. 

We derive the corresponding values of  $\Ml'$, $\Mm'$, $M_0$, and $\al$ in our simulations that closely reproduce the best fit number density of central and satellite galaxies provided by \citet{Zheng07} so that the relative strength of 1-halo and 2-halo terms is consistent with the observation despite the slight discrepancy in the mass function. In detail, we account for the difference in the mass function by rescaling $\Mm'$ and $M_0$ {\it up} by $5 \sim 6\%$, as our halo mass scale in this range is overestimated by $\sim 5\%$, while adjusting $\Ml'$ or $\al$. Note that we do not fit to the observed correlation function directly. We show one example of such fits (we call `Model LRG') and the resulting correlation function (Table \ref{tab:tHODz0.3} and Figure \ref{fig:Corallz0.3M}). As shown in Table \ref{tab:tHODz0.3}, $\Mm'$ and $M_0$ for Model LRG are increased by $6\%$ relative to the Zheng's values, while $\sigma_M$, $\Ml'$, and $\al$ remain nearly the same as Zheng's.

In Figure \ref{fig:Corallz0.3M}, the clustering of Model LRG is fairly consistent with the observed clustering in real and redshift space except for $r < 0.5\hMpc$: the slope of $\xi$ over $r < 0.5\hMpc$ is determined by our smallest bin at $r \sim 0.2 \hMpc$ that is likely subject to the effect of our force resolution. In addition, a small discrepancy remains near $r\sim 2.5\hMpc$ in real space and $r\sim 1-2 \hMpc$ in redshift space. Again, note that Model LRG is derived based on the fitted values of central and satellite number densities from \citet{Zheng07} rather than a direct fit to the observed clustering data. Although a slight modification in $\mng$ of Model LRG may bring the resulting $\xi$ closer to the observed $\xi$, we find this unnecessary as we will focus on quantitative but low-precision comparisons.

\begin{figure*}[t]
\epsscale{1.6}
\figurenum{10}
\plotone{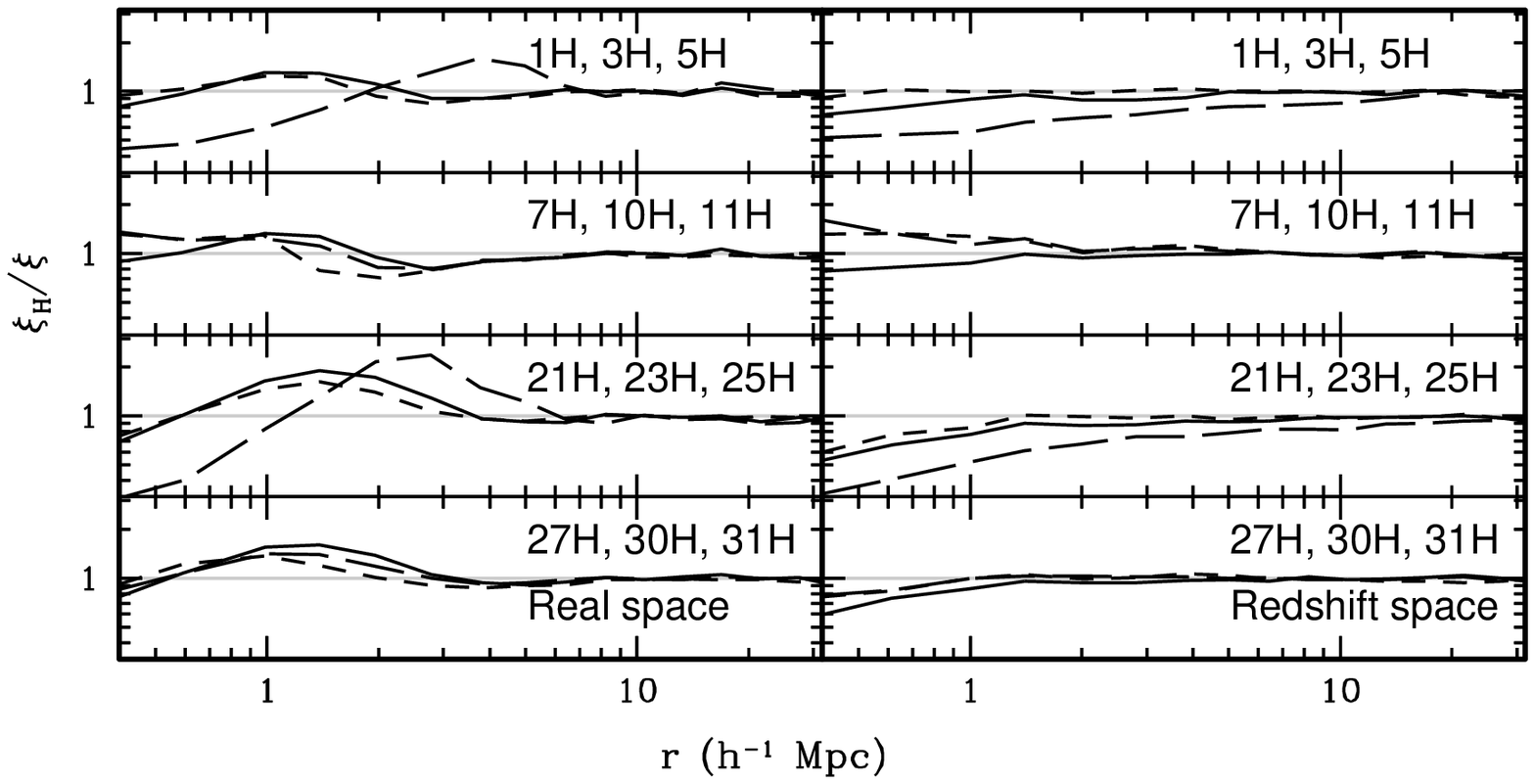}
\caption{The comparisons between $\xi$ at $z=0.3$ for passively flowing galaxies (Models 1--31) and $\xi_{\rm H}$ for prompt populations (Models 1H--31H) that are assigned at $z=0.3$ using $\mng$ and the second moment of Models 1--31. Left : the ratio of correlation functions in real space. Right : redshift space. Solid, dashed, and dotted lines are for three different models, for example, solid line in the top panels are the ratio of correlation functions of Model 1H to Model 1, long-dashed line : Model 3H to Model 3, and short-dashed line : Model 5H to Model 5. The clustering of the two cases is indistinguishable on large scales while different on small scales.  }
\label{fig:Corallz0.3H}
\epsscale{1}
\end{figure*}

Model LRG produces a shape of $\xi$ which is roughly a power law, mainly because of the existence of the shoulder in HOD that drives a smaller 1-halo term relative to the 1-halo term of dark matter. The shoulder implies a smaller fraction of satellites that is mostly from massive halos in this case, and so smaller pair counts relative to that of dark matter \citep[see][]{Seljak00,Berlind02}. Model 11 in Figure \ref{fig:Corallz0.3M} also shows a shoulder in HOD. Despite the small overall satellite fraction, pair counts of this model imply a Poisson distribution for $M> 10^{14}\Msun$ unlike Model LRG (\S~\ref{sec:z1Mz2MNN}). Nevertheless, the correlation function does not deviate too much from a power-law shape. We revisit this and relate this to the effect of evolution in \S~\ref{sec:z0.3H}.

We have found that the observed LRG clustering is mostly reproducible in our simulations with large values of $\Ml/\Mm$ and $\al$ imposed directly at $z=0.3$, as motivated by \citet{Zheng07}. Meanwhile, the observed correlation function of the LRGs does not eliminate the possibility of the LRGs having flowed passively from $z \sim 1$: from Figure \ref{fig:Corallz0.3M}, those near Model 10 or Model 11 will not fare much worse with the observed LRG clustering than models near Model LRG would.

However, if we take the fit to the LRGs by \citet{Zheng07} as concrete, the shoulder in their HOD at $z=0.3$ requires some amount of non-passive flow evolution, as it is difficult to reproduce a large $\Ml/\Mm$ and $\al$ at $z=0.3$ with passive flow. In general, we need a process to suppress the LRG satellites in low or intermediate mass halos (i.e., $10^{14}\Msun-10^{15}\Msun$) while we need more LRGs in very massive halos. As discussed in \S~\ref{subsec:z1M}, including dynamical friction will cause our LRG progenitors to merge to the center, increasing $\Ml/\Mm$ and possibly moving $\al$ around. This process will be more efficient in lower mass halos, as lower mass halos are in general older than more massive halos, and so they would have had more time for this destructive process \citep{Krav99,Taffoni03,Zentner03,Zentner05,vanB05b,Taylor05}.
For very massive halos, however, we face a contradiction of requiring more satellites at $z=0.3$ than the conserved number density can give, which means we need a new source of LRGs in these very massive halos. These new LRGs, which are admittedly rare, must have built up enough mass between $z=1$ and $z=0.3$ through dry merging preferentially in very massive halos. Note however that some models with large initial $\al$ ($\sim 2$) do not violate the number conservation for high mass halos. For example, Model 9\nfo\ (\S~\ref{subsec:z1n4}) might reduce to a large $\Ml/\Mm$ and $\al$ after non-passive evolution without a necessity for new satellites for very massive halos.

\section{Galaxies assigned at $ z=0.3$ : finding a hidden signature of passive flow evolution} \label{sec:z0.3H}
We have assigned galaxies based on the assertion that the number of galaxies is only a function of halo mass at the initial redshift. However, the subsequent evolution may cause a deviation from this assertion; the clustering or the HOD at $z=0.3$ may no longer be a function of halo mass alone. Several studies \citep{NFW97,Bullock01,Wech02,Sheth04,Gao05,Wech05,Zhu06,Harker06,Croton06,Wetzel07,Jing07,Gao07} have indicated that details of halo formation history can affect clustering and properties of halos of the same final mass. For example, at a given final mass, halos that formed earlier show stronger concentration, reflecting the high density of the Universe at early times, and contain smaller number of satellites due to ongoing dynamical friction. It is also shown that, for low mass halos, the clustering of early forming halos at a given final halo mass is stronger than late forming halos \citep{Gao05,Wech05,Zhu06,Harker06,Croton06,Wetzel07,Jing07,Gao07}. The clustering trend weakens and reverses for very massive halos, in which most of the galaxies studied in this paper reside.

In this section, we look for signatures of passive flow evolution in correlation function at $z=0.3$ that cannot be parameterized by halo mass at $z=0.3$ alone. As a caveat, as we assigned galaxies to halos only based on halo mass at the initial redshift of $z=1$ or 2, we ignored the effect of halo assembly history on galaxy occupation prior to the initial redshift. It is important to note that the condition of passive flow may draw evolutionary signatures that are different from non-passive populations. We construct `prompt' populations at $z=0.3$ that have the same $\mng$ and $\NNs$ as those from passively flowing galaxies but that are randomly distributed among halos of a given mass. As a result, this `prompt' population is a galaxy population assigned only as a function of halo mass at $z=0.3$, i.e., without any evolutionary effect. We compare the clustering of passively flowing galaxies, i.e., evolved populations, with that of the `prompt' populations at $z=0.3$.

\begin{figure}[b]
\epsscale{0.8}
\figurenum{11}
\plotone{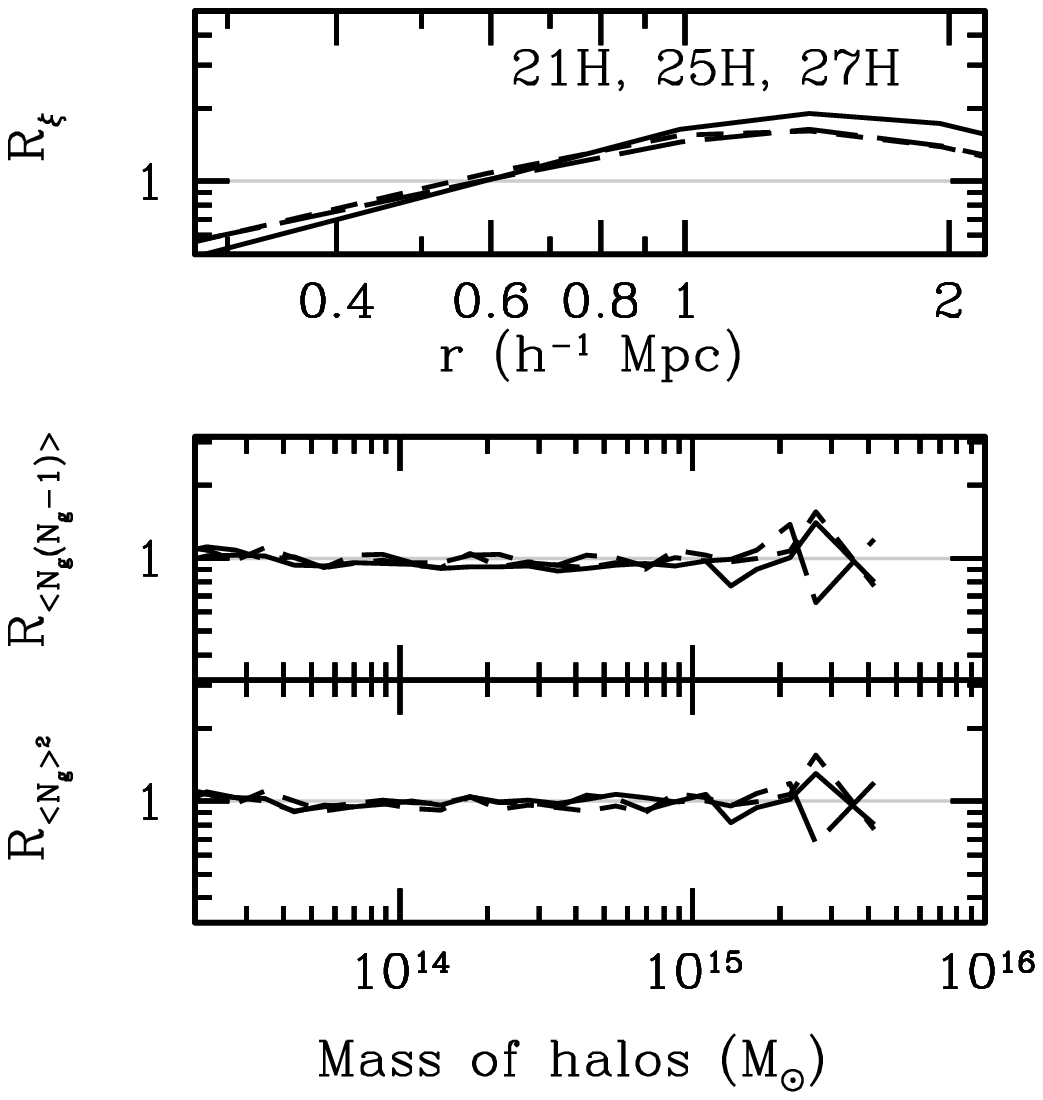}
\caption{The comparisons of pair counts at $z=0.3$ between galaxies passively flowing from $z=2$ and prompt galaxy populations that are directly assigned at $z=0.3$ (Models 21H (solid), 25H (long-dashed), and 27H (short-dashed). Top : ratios of correlation functions of Models 21H, 25H, and 27H to Models 21, 25, and 27, respectively. Middle : ratios of $\NNs$. Bottom : ratios of $\mng^2$. The difference in $\xi$ is probably due to the difference in a radial distribution of galaxies within a halo, as the first and the second moments of HODs are almost identical for the evolved and the prompt populations.}
\label{fig:z0.3HNN_N}
\end{figure}

In detail, we calculate $\mng$ by locating passively flowing galaxies (\zoM\ and \ztM) in the halos at $z=0.3$ and consider the first galaxy in each halo as the central galaxy and the rest as satellites. We then reassign galaxies back to random halos at $z=0.3$ using the derived $\mng$ for central and satellite galaxies. In \S~\ref{sec:z1Mz2MNN}, we claimed that these passively flowing galaxies at $z=0.3$ are consistent with central galaxies in a nearest integer distribution and satellite galaxies in a Poisson distribution. We therefore assume these statistics to construct the `prompt' populations. The central galaxies are located at the most bound particles of halos, while the satellite galaxies trace mass. We label these models as \zHzoM\ and \zHztM, where \zHzoM\ are from \zoM, and \zHztM\ are from \ztM. Some models result in a number of galaxies without any associated host halos (about $0.06 \sim 0.2 \%$) at $z=0.3$, especially from \zoM. We simply ignore these galaxies.

\begin{figure}[t]
\epsscale{0.8}
\figurenum{12}
\plotone{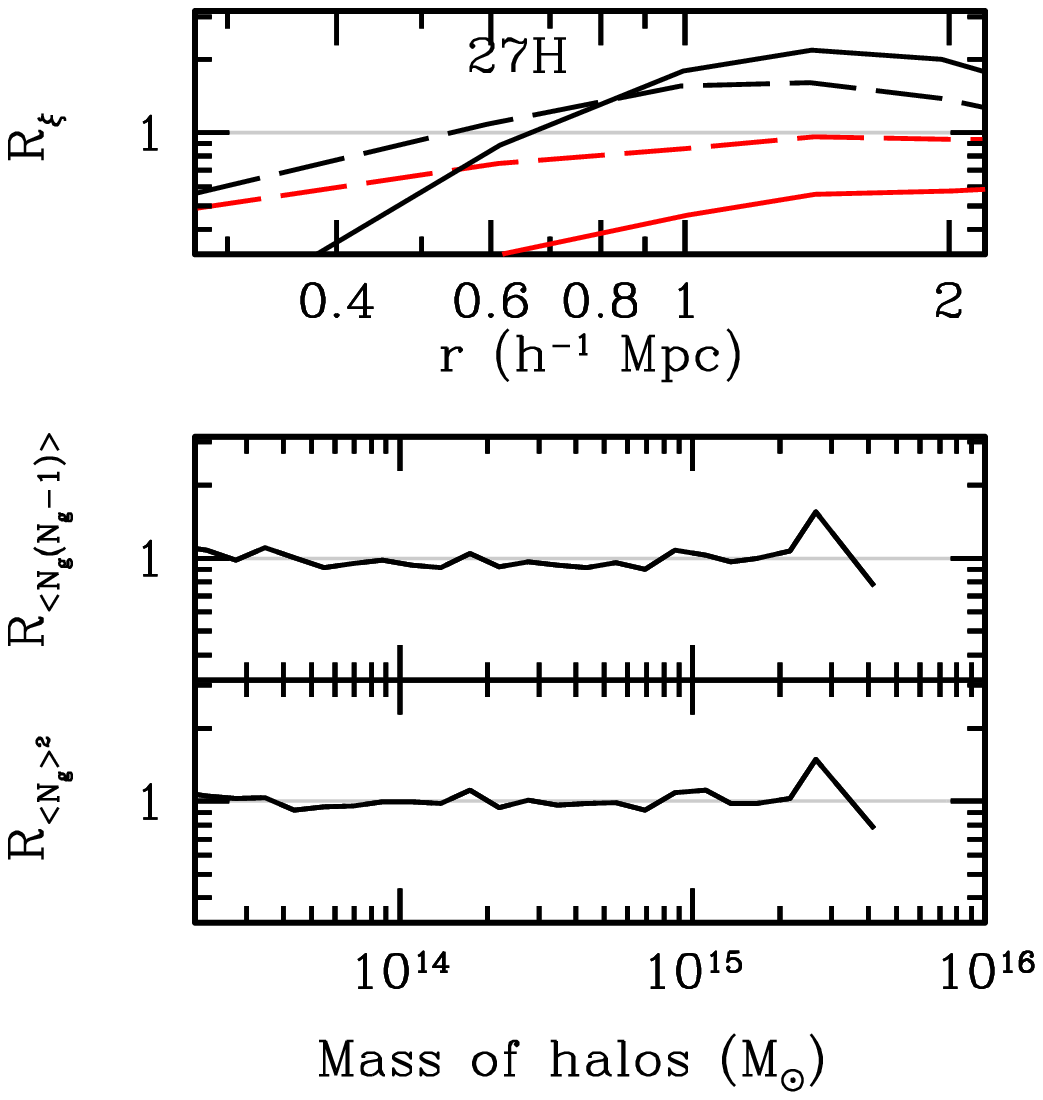}
\caption{The comparisons between Model 27 and a prompt population 27H when the central galaxies of Model 27H are displaced from the center of a halo (solid lines), compared with the original Model 27H (dashed lines). In the top panel, black lines are for the real-space correlation functions and the red lines are for the redshift-space correlation functions. There is no difference in $\NNs$ and $\mng^2$ between the adjusted Model 27H and the original Model 27H. }
\label{fig:z0.3Hnocen}
\end{figure}

\begin{figure*}
\epsscale{1.9}
\figurenum{13}
\plotone{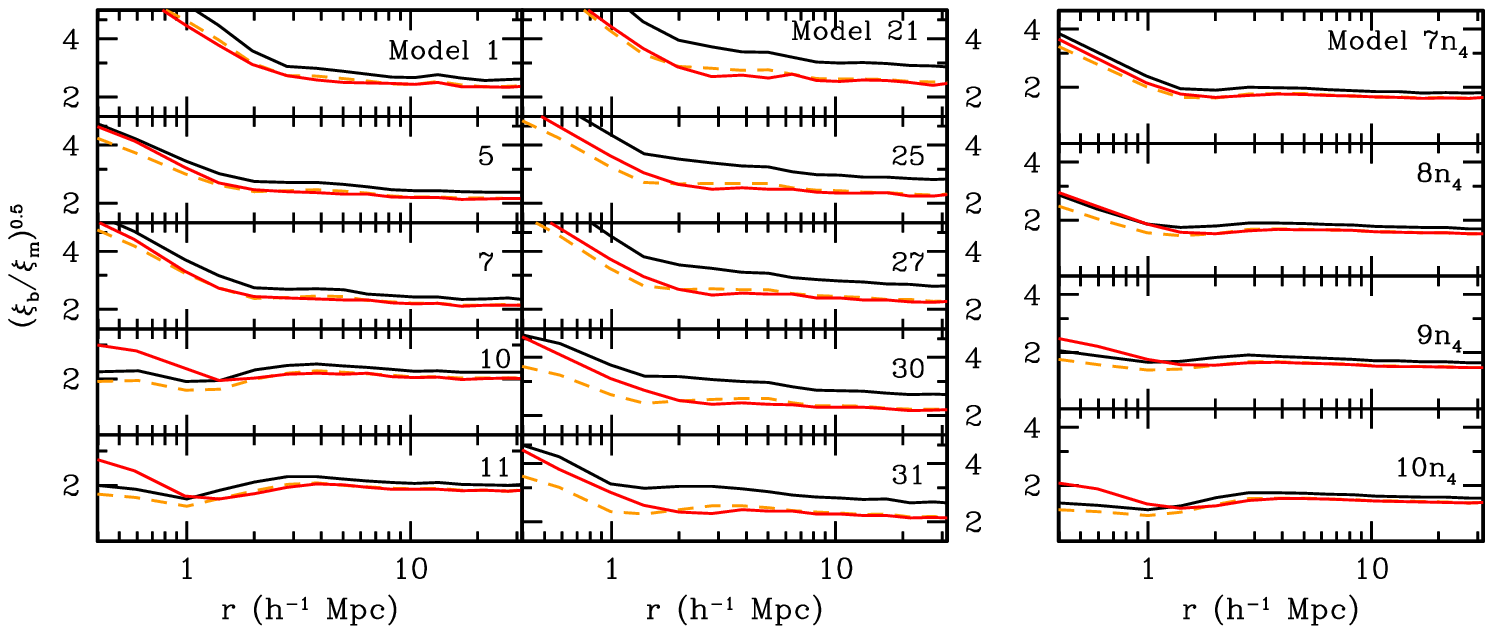}
\caption{The evolution of bias. Left: bias evolution for Models 1, 5, 7, 10, and 11 at $z=0.6$ (black lines) and $z=0.3$ (red). Middle: bias evolution for Models 21, 25, 27, 30,31 at $z=1$ (black lines) and $z=0.3$ (red). Right: bias evolution for \zoMfn\ at $z=0.6$ (black lines) and $z=0.3$ (red). Orange lines denote the expected bias values at $z=0.3$ assuming linear bias evolution adopting a scale-dependent bias $b(r)$ and a growth factor $G(r)$ from $z=0.6$ (left and right) or $z=1$ (middle). Y-axes of all panels are in logarithmic scale.}
\label{fig:biasevol}
\epsscale{1}
\end{figure*}

Figure \ref{fig:Corallz0.3H} shows the difference in clustering between the evolved populations (\zoM\ and \ztM) and the prompt populations at $z=0.3$ (\zHzoM\ and \zHztM). Both populations produce very similar large-scale clustering for the given $\mng$, which means that it is a good approximation in this scale range to take the halo mass as the only variable to decide the number of galaxies per halo, at least for populations in very massive halos, such as LRGs. The difference remains only on small scales, and is larger for models constructed at $z=2$ than at $z=1$. In detail, the clustering of \zHztM\ is larger than \ztM\ over $r=1-2 \hMpc$, even by up to a factor of two for the models with a large number of initial satellites (Models 21--23). 

In Figure \ref{fig:NNsz2all}, we have seen the consistency between the small-scale clustering and the corresponding pair counts among different models. Similarly we will look for any remaining difference in the second moments between the evolved populations and the prompt populations, that is, the effect of any deviation from a Poisson distribution for the evolved satellite populations. Figure \ref{fig:z0.3HNN_N} illustrates the ratios of  $\mng^2$ and $\NNs$ between Models 21H, 25H, 27H and Models 21, 25, and 27, compared with the ratio of $\xi$. The ratio of ${\NNs}$ shows that the evolved populations at $z=0.3$ are statistically very similar to the prompt populations. The noisy fluctuation at the most massive halos contributes only little to the overall clustering. As a caveat, pair counts in intermediate mass halos with a small satellite fraction are not very sensitive to the statistics because they are dominated by central galaxy-satellite galaxy pairs, which in turn depends on $\mngc$ and $\mngs$. Because we set these quantities to be almost identical, the shape of $\NNs$ is naturally very similar between the passive and prompt populations as long as satellite populations asymptotically converge to Poisson statistics. Based on the similarity of the resulting pair counts, therefore, the small-scale difference observed in $\xi$ appears rooted in the evolutionary effect of the halo profiles rather than pair counts. 

From Figure \ref{fig:Corallz0.3H}, the difference in $\xi$ is not constant over scale: the ratio decreases below 1 for $r<1\hMpc$ but increases above 1 near $r \sim 1\hMpc$. The shape of the non-monotonic trend implies a broadened $\xi$ for the prompt populations (except for Models 9, 10, and 11), more close pairs in the evolved populations, and therefore the evolved galaxies taking a steeper radial gradient than the prompt populations whose satellite galaxies follow dark matter gradient at $z=0.3$. For comparisons, the cluster galaxy radial gradient suggested from non-passive studies as well as observations is close to or slightly shallower than mass profile \citep[e.g.,][and observational references therein]{Diemand04,Gao04a,vanB05a,Nagai05,Weinberg06} with variations in morphological type of galaxies \citep{Springel01}. We try below to provide some physical intuition for why the galaxy gradient would be steeper than the mass gradient for the passively flowing populations.

Our LRG progenitors in passive flow resided in massive halos that correspond to high and rare density peaks at early times. At low redshift, material from these high peaks tends to be more centrally concentrated within a host halo than that from lower density peaks and the overall mass distribution \citep{Diemand05,Moore06}. Mass coming from these lower density regions remains with few galaxies, as we do not allow any new sources of galaxies. As a result, a host halo at low redshift will have a high galaxy density in its inner region and little galaxy content accreted into the outer region of the halo. In other words, the memory of the typical host-halo size at high redshift remains in the process of passive flow, resulting in the small-scale difference in $\xi$. On the other hand, the trend is reversed for Models 9--11 in Figure \ref{fig:Corallz0.3H}, where the radial gradient of galaxies appears weaker for evolved population, that is, galaxies are on average less concentrated than mass in this case. These models started with no satellites and probably did not have enough time to populate satellites at a close distance while Models 29, 30, and 31 were able to. In summary, passive flow will result in the galaxies being more concentrated than dark matter within a host halo, unless these galaxies are recent descendants of central galaxies. 

We found that for some cases the difference in small-scale clustering due to passive flow evolution is enough to cause the correlation functions of evolved galaxies to appear more like a power law compared to the prompt populations, and vice versa. For example, Models 7H, 10H, and 11H will give much worse fits to the observed LRG points than Models 7, 10, and 11 did in Figure \ref{fig:Corallz0.3M}. Therefore the effect of evolution could alter the best fit HOD parameters to the observed clustering. Note that Model LRG in \S~\ref{sec:z0.3M} corresponds to a prompt population.

The stronger radial gradient for evolved populations could also be interpreted in the context of the concentration evolution of the halos themselves with redshift \citep{Bullock01,Wech02,Zhao03a,Zhao03b}, rather than in the context of a radial stratification within a halo with redshift. In other words, what we observe could be due to the passively flowing galaxies at low redshift and at a given final halo mass being more likely distributed among older halos, where mass concentration is on average stronger \citep{Bullock01,Wech02}, as they are halos from massive progenitors at $z=2$ or $z=1$. As the host halos are fairly massive, it is likely that the corresponding difference in large-scale clustering is little, as in Figure \ref{fig:Corallz0.3H} \citep{Gao05,Wech05,Harker06,Wetzel07,Jing07,Gao07}.

In redshift space, the difference generally appears as a relative suppression in the prompt populations, except for Models 9, 10, and 11, especially for $r< 1\hMpc$. The finger-of-God effect evacuates more small-distance pairs in the prompt populations than in evolved populations. As the pairwise velocity dispersion within virialized halos increases with distance \citep{Sheth01}, the evolved populations with closer pairs will have a weaker finger-of-God effect. Also, according to \citet{Diemand05}, the velocity dispersion of material from the rarer peaks is lower than that of matter at a given radius from halos, which may contribute to the weaker finger-of-God effect as well.

We consider the possibility that central galaxies in the passively flowing populations have not settled down at the center of their host halos at $z=0.3$ after a series of halo merging events. This will also change the small-scale clustering relative to \zHzoM\ and \zHztM\ but probably in an opposite direction to what we have observed. We test an extreme case in which no halos in prompt populations host central galaxies. In detail, we substitute central galaxies in Model 27H with random satellite galaxies. Figure \ref{fig:z0.3Hnocen} shows that removing the central galaxy strengthens the difference between Model 27H and Model 27. In redshift space, the effect is more intense: the finger-of-God effect is strongly enhanced up to $r \sim 10 \hMpc$ due to the missing pairs at small separation. This implies that the evolved populations at $z=0.3$ probably have a relatively well-positioned central galaxy and, again, satellites closer to the center.

To summarize, the large-scale clustering of populations that have evolved through passive flow is reproducible with $\mng$, and therefore we do not observe significant environmental effects on large-scale clustering in this sample. The satellite galaxies of passively flowing populations closely follow a Poisson distribution, and so they are indistinguishable in pair counts from the prompt populations with the same $\mng$.  The effect of passive flow evolution, however, appears in the spatial distribution of galaxies within a halo, in that the passively flowing populations, unless they are recent descendants of central galaxies, show on average more centrally concentrated distribution than the prompt populations (i.e.,  mass profile).

\section{Evolution of bias}\label{sec:BiasRed}

In this section, we study the evolution of bias of passively flowing galaxies and compare it to the linear theory for passive flow evolution.

\begin{figure*}
\epsscale{1.7}
\figurenum{14}
\plotone{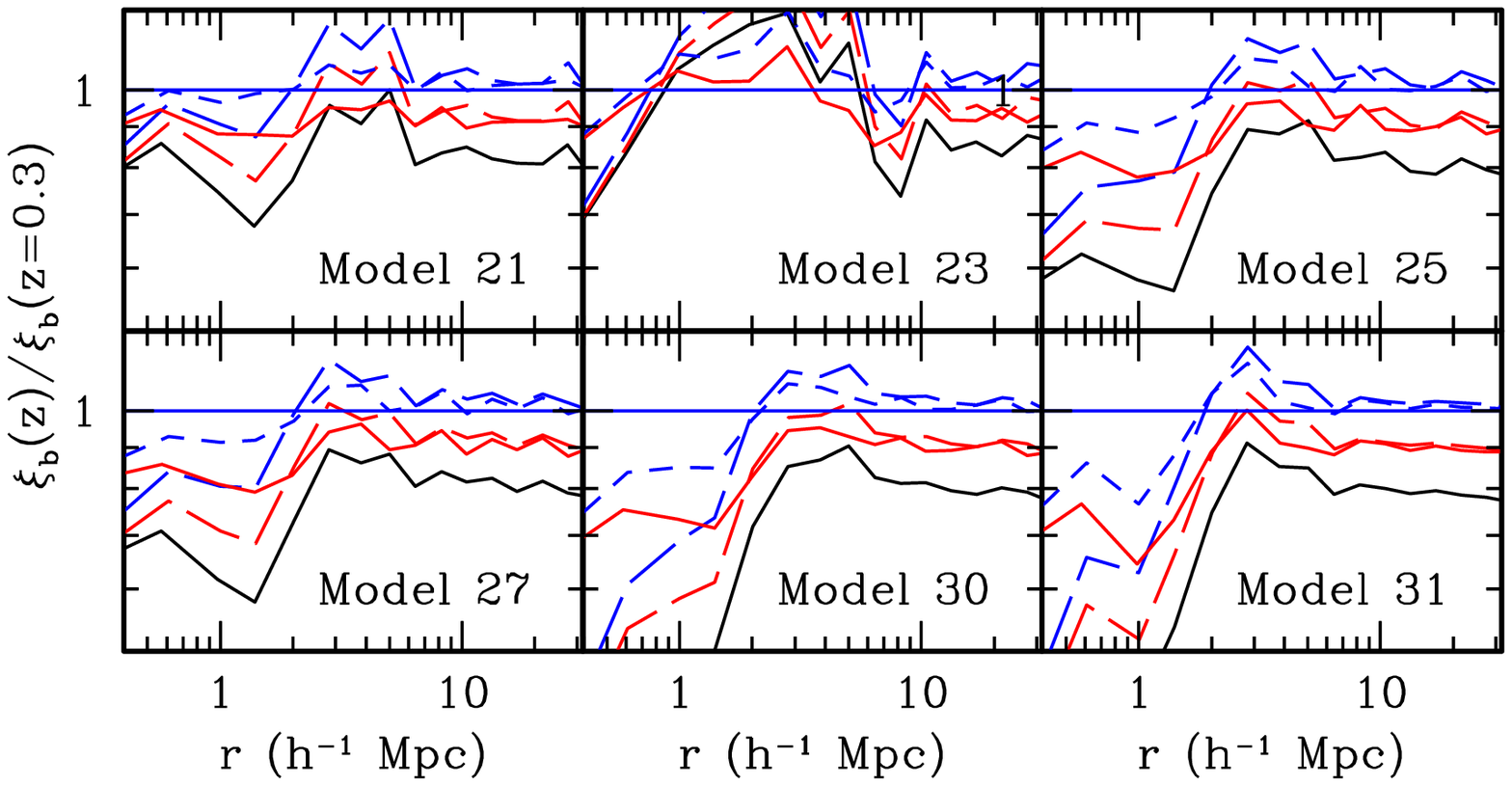}
\caption{The evolution of real-space correlation function of passively flowing galaxies for Models 21, 23, 25, 27, 30, and 31. Correlation functions are divided by the correlation function at $z=0.3$ (solid lines). Black lines : $z=1$. Red : $z=0.6$. Blue : $z=0.3$. Long-dashed lines : the expected correlation functions at $z=0.6$ (red) and at $z=0.3$ (blue) assuming the linear bias evolution from $z=1$ (as defined in \S~\ref{sec:BiasRed}). Short-dashed : the expected correlation function at $z=0.3$ assuming the linear bias evolution from $z=0.6$. One finds that growth of clustering is relatively suppressed near $r=2-5 \hMpc$, compared to the growth on other scales. }
\label{fig:evolcorrz2M}
\end{figure*}

Figure \ref{fig:biasevol} shows the typical bias evolution for \zoM\ and \ztM. As the exclusion effect in a halo finder is prevailing at initial redshifts, the evolution of bias is considered only between $z=0.6$ and $z=0.3$ for \zoM\ (left panels) and between $z=1$ and $z=0.3$ for \ztM\ (middle panels). The bias factor at $z=0.3$ is compared to the expected linear bias evolution \citep{Fry96,Tegmark98} where we define `linear bias evolution' by unconventionally adopting a scale-dependent bias $b(r)$ and a scale-dependent growth factor of dark matter $G(r)$ at $z=0.6$ (left, for \zoM) or $z=1$ (middle, for \ztM). That is, we define the growth factor $G(r)$ at given redshift $z$ as $(\xi_{m,z}(r)/\xi_{m,z_0}(r))^{1/2}$ where $z_0=0.6$ for \zoM\ and 1 for \ztM, and the bias factor $b_{z_0}(r)$ as $(\xi_g(r)/\xi_m(r))^{1/2}$ at $z_0$. Then the `linear bias evolution' at $z$ expected from $z_0$ is derived from

\begin{equation}
b_{\rm lin}(r)=(b_{z_0}(r)-1)/G(r)+1.
\end{equation}
We then compare the expected bias, $b_{\rm lin}(r)$, with the actual bias, $b(r)=(\xi_g(r)/\xi_m(r))^{1/2}$ measured at $z$.

The bias values at $z=0.3$ are consistent with the expected bias evolution over $r > 2 \hMpc$ in most of the models, with only small discrepancies on smaller scales. In detail, the bias at $z=0.3$ implies that the growth in clustering becomes slightly impeded over $r=2-6 \hMpc$ but enhanced over $r < 2 \hMpc$ either relative to the growth of mass or relative to the linear bias evolution. The discrepancies appear bigger for \ztM, probably due to the longer evolution time and the larger initial bias.

For \zoMfn\ (right panel), the decrease in bias is less than that of Models 7--10 due to the smaller initial bias values and is also consistent with the expected linear bias evolution over $r > 2 \hMpc$. Due to the dominance of lower mass halos in these models, their bias factor at low redshift deviates from the scale-independent bias on smaller scales than Models 7--10. 

To summarize, the evolution of bias for the passively flowing galaxies is relatively `linear' for $r > 2 \hMpc$ although slightly scale-dependent.

\begin{figure*}
\epsscale{1.2}
\figurenum{15}
\plotone{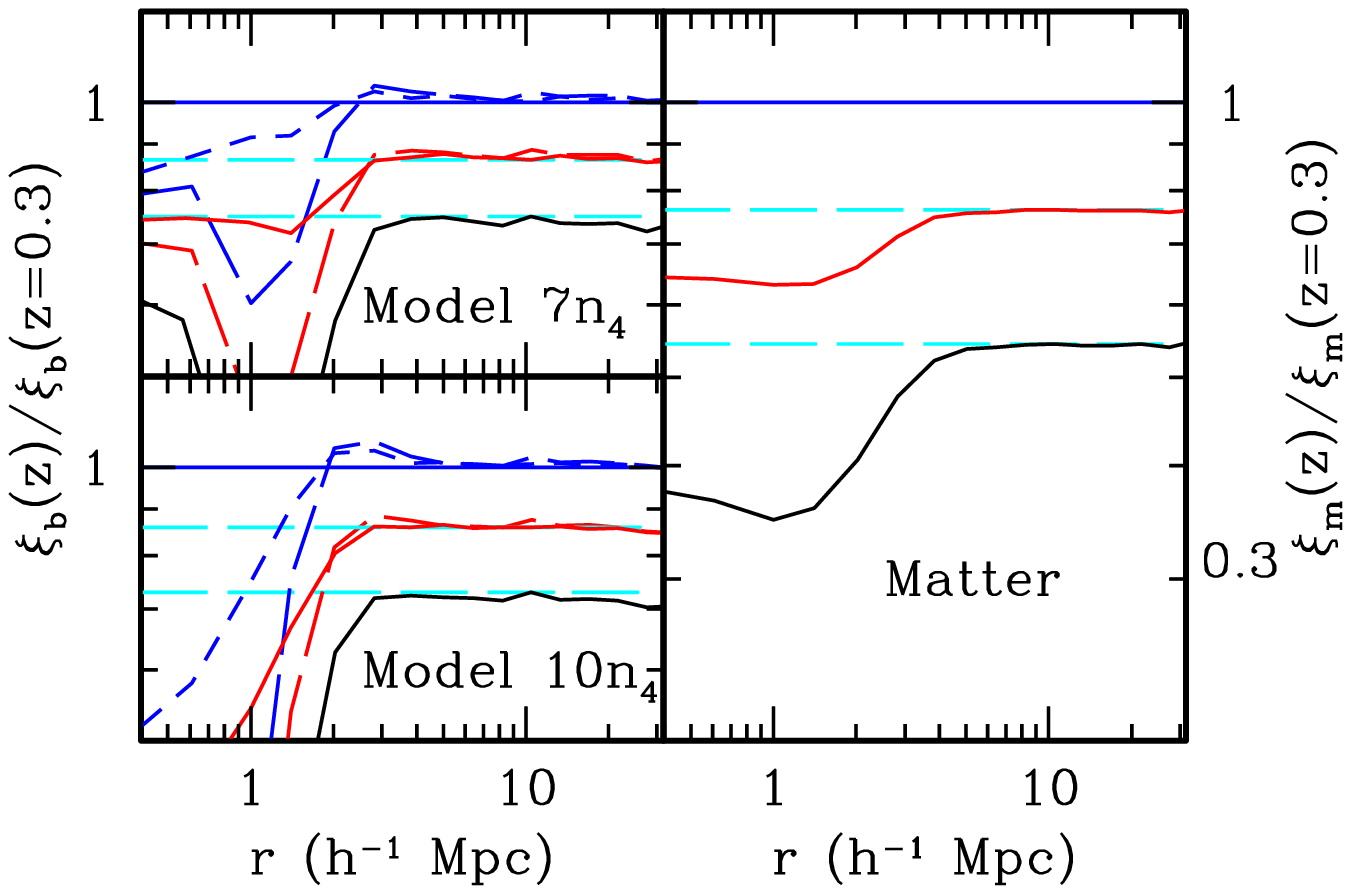}
\caption{The evolution of correlation functions of passively flowing galaxies for Models 7\nfo\ and 10\nfo ($\xi_b$, left panel) and matter ($\xi_m$, right). Correlation functions are divided by the correlation function at $z=0.3$ (solid lines). Black lines : $z=1$. Red : $z=0.6$. Blue : $z=0.3$. Long-dashed lines : the expected correlation functions at $z=0.6$ (red) and $z=0.3$ (blue) assuming the linear bias evolution from $z=1$. Short-dashed : the expected correlation functions at $z=0.3$ assuming the linear bias evolution from $z=0.6$. Cyan lines are at constant values and are drawn to help to estimate a scale-dependence of the growth in correlation functions. One finds that the suppression of growth near $r=2-5 \hMpc$ is not apparent at all for dark matter or galaxies with small bias ($b<2$).} 
\label{fig:evolcorrz1n4}
\end{figure*}

\section{Evolution of correlation function: a signature of infall?} 
\label{sec:evolCor}
In this section, we attempt to characterize the evolution of clustering of passively flowing galaxies on quasilinear scales.

In Figure \ref{fig:evolcorrz2M}, real-space correlation functions at $z=1$, 0.6, and 0.3 for a number of models among \ztM\ are divided by a real-space correlation function at $z=0.3$. The apparent features are the suppression of growth near $r=2-5 \hMpc$, and the large growth inside the radius (also see the left panels of Figure \ref{fig:fHODz1Mall} and \ref{fig:fHODz2Mall}). The scale where the suppression appears is slightly larger than the transition from a 1-halo to a 2-halo term where the inflection of a power-law model of biased correlation functions occurs. We compare the growth between $z=1$ and $z=0.6$ with what is expected from the linear bias evolution (as defined in \S~\ref{sec:BiasRed}) from $z=1$: the growth between $z=1$ and $z=0.3$ and between $z=0.6$ and $z=0.3$ is compared with the linear bias evolution from $z=1$ and $z=0.6$, respectively. The figure shows that the suppression in growth is not explained by linear evolution of clustering of biased tracers.

This effect may be interpreted as pairs at $r=2-5 \hMpc$ moving rapidly to smaller scales as we go to lower redshift, that is, an evacuation of pairs on this scale as structure grows linearly on larger scales but nonlinearly on smaller scales.
The feature is less obvious and appears at a larger radius for \zoM.

In Figure \ref{fig:evolcorrz1n4}, we test whether we see the corresponding feature in the different populations or in the matter correlation function. The evolution of matter correlation function does not show an obvious suppression, probably because matter is in steady state in flow from a linear to nonlinear region of radius. On the other hand, 1-halo and 2-halo terms of the galaxies do not equally weight the corresponding terms for matter \citep{Schulz06} but trace the flow of matter differently in the two regions, breaking the steady state. 

To examine whether the suppression is a prevalent behavior of biased tracers, we test \zoMfn. Figure \ref{fig:evolcorrz1n4} shows the suppression in growth of correlation function is not obvious at all for \zoMfn\, and so the growth of clustering is scale-independent for $r > 2-3\hMpc$. As a minor point, when compared to the matter correlation functions (right panel), the nonlinear growth of the correlation functions for \zoMfn\ happens on smaller scales than that of matter; this slightly overpredicts the linear bias evolution at $r=2-4 \hMpc$ beyond a scale-independent growth (in the left panels). The missing feature of suppression for \zoMfn\ therefore implies that the feature appears preferentially in strongly biased tracers ($b \gtrsim 2$ at $z=0.3$).

\section{Conclusions}
We have used dissipationless N-body simulations to study the effect of passive flow evolution on galaxy clustering and halo occupation distributions. We assumed populations of progenitor galaxies at $z=1$ and $z=2$ with a wide range of initial HODs, and then studied their properties as they flowed passively to $z=0.3$. We investigated for a region of the parameter space at low redshift constrained by passive flow, especially in the halo occupation distribution and in galaxy clustering. Our results are summarized as follows.

Passive flow results in an asymptotic convergence in halo occupation distributions and galaxy clustering. The distribution of the average number of galaxies per given halo mass converges toward a power law for a broad range of the initial halo occupation distributions. The values of $\Ml/\Mm$ decreases with time, except for the cases with very small initial $\Ml/\Mm$. The value of $\al$ asymptotically converges toward unity. Both evolutionary behaviors result in $\mng$ without a shoulder and thus a shape close to a power law at low redshift.

A similar convergence is observed in the evolution of correlation function. While it is not surprising to find that the large-scale convergence is consistent with the expected clustering from linear bias evolution \citep{Fry96,Tegmark98}, the intermediate-scale clustering also shows a fair degree of convergence in its shape and amplitude when the galaxies have evolved from high redshift ($z\sim 2$). The remaining differences in small-scale clustering appears consistent with the differences in $\mng$. Modeling these intermediate and small scales is left for future research.

The asymptotic convergence is enhanced if the galaxies passively flow from higher redshift and so have more time to flow. We also find that the satellite populations are fairly large at low redshift as a result of passive flow evolution.

We have tried the same tests with galaxy populations of a larger number density and found qualitatively similar results. The asymptotic convergence in the HODs and correlation functions is not an extreme behavior for the very high mass halos but is a general result of passive flow evolution. We conclude that the convergence of $\Ml/\Mm$ and $\al$ is the result of galaxies asymptotically becoming less biased with respect to mass with time.

Passive flow evolution drives satellite galaxies to converge toward the Poisson distribution. Other studies for non-passive evolution reproduce the Poisson distribution as well, implying that the merging, formation, and destruction of galaxies happen so that the resulting populations remain in the Poisson distribution.

The comparison of these characteristics of passive flow evolution with the observed LRG data hints at any non-passive flow processes during the evolution of the LRGs. A suggested existence of a shoulder between $\Ml$ and $\Mm$ or a large $\al$ in the HOD of LRGs when fitted to the observed clustering data \citep[][also see \citet{Blake07}, but see  \citet{Kulkarni07} and \citet{Ho07} for different results]{Zheng07} implies that LRGs have not undergone a strict passive flow evolution. 
These discrepancies could be due to dry galaxy merging between two LRGs or new LRGs arising between the initial and the final redshifts, with different efficiencies in different environments. The discrepancies should show self-consistent consequences in other observational properties such as in the luminosity function of the LRGs \citep[e.g.,][]{Wake06,Brown07}. See \citet{White07} for an example of such a study.

We compared two populations with the identical $\mng$ at $z=0.3$, one that is reached after passive flow evolution and the other constructed from random halos at $z=0.3$, to study whether the evolution imposes any distinct signature in the second moments of the HOD and clustering.  The evolved population and the prompt population showed no significant difference in the second moments of the HOD when both satellite populations are assumed to be Poisson at the initial redshift. The effect of evolution did not impose a noticeable environmental dependence in large-scale clustering. In small-scale clustering, inside halos, we find that the spatial distribution of the passively flowing galaxies is on average more centrally concentrated than mass.

The evolution of bias for passively flowing galaxies were consistent with linear bias evolution on quasilinear as well as large scales once the scale-dependent growth factor of dark matter is considered.

We identified a suppression of growth of galaxy clustering near $r=2-5\hMpc$. The feature is observed in a population with a large initial bias, but not in a population with a small bias nor in clustering of matter. 

An interesting extension of this research will be to investigate the characteristics of passive flow evolution in velocity space. Once we understand both spatial and velocity signatures of passively flowing galaxies, we will have a better handle to model the clustering evolution of such galaxy populations. We will also extend our work to non-passively flowing galaxies, using phenomenological galaxy merging prescriptions.
\acknowledgements
We thank Zheng Zheng for providing his best-fit HOD parameter estimates for the LRGs and his valuable comments on this paper. HS, IZ, and DJE were supported by grant AST-0407200 from the National Science Foundation.

\clearpage

\onecolumn

\begin{deluxetable}{l|ccccc|ccccc}
\tablewidth{0pt}
\tabletypesize{\small}
\tablecaption{\label{tab:tHODz1} The initial and final HOD parameters of the galaxies passively flowing from $z=1$ }
\startdata\hline\hline
Model&$\al$ & $\Ml/\Mm$&$\Mlft$\tablenotemark{a}&$\Mmft$\tablenotemark{a}&$\fcen$&$\al$ & $\Ml/\Mm$&$\Mlft$\tablenotemark{a}&$\Mmft$\tablenotemark{a}&$\fcen$\\
    &           && $z=1$  &       &      &       &       &$z=0.3$&     & \\\hline
1   &0.5& 2.061&1.0   &0.4853 &0.58  &0.829& 2.149&1.847  &0.8592  &0.53\\
2   &1& 2.167  &1.0   &0.4614 &0.63  &1.06&2.414  &1.964  &0.8133  &0.56\\
3   &2& 1.929  &1.0   &0.5185 &0.53  &1.67& 2.159 &1.994  &0.9236  &0.48\\
4   &0.5& 9.814&4.0   &0.4076 &0.74  &0.916& 4.223 &3.00   &0.7097  &0.67\\
5   &1& 11.16  &4.0   &0.3585 &0.88  &1.11& 5.979 &3.704  &0.6195  &0.79\\
6   &2& 11.90  &4.0   &0.3360 &0.97  &1.23& 7.408 &4.281  &0.5779  &0.86\\
7   &0.5& 26.50&10    &0.3774 &0.83  &0.927& 5.390&3.525  &0.6540  &0.74\\
8   &1& 29.43  &10    &0.3398 &0.95  &1.033& 7.353&4.302  &0.5851  &0.84\\
9   &2& 30.39  &10    &0.3291 &0.99  &1.10& 8.646 &4.890  &0.5655  &0.88\\
10  & &        &      &0.3275 &1     &1.04& 9.052 &5.091  &0.5625  &0.88\\
11  & &        &      &0.2141 &1     &1.43& 12.01 &10.39  &0.8651  &0.91\\
\enddata
\tablenotetext{a}{Mass of halos are in the unit of $10^{14} \Msun$.}
\tablecomments{We show the input HOD parameters at $z=1$ and the best fit HOD parameters at $z=0.3$ for \zoM.  Note that $\Ml$ and $\al$ at $z=0.3$ are fitted over $M>\Ml$ in order to better describe the shape of satellite HODs at the massive end. The value $\fcen$ is the fraction of central galaxies to the total number of galaxies. }
\end{deluxetable}

\begin{deluxetable}{l|ccccc|ccccc}
\tablewidth{0pt}
\tabletypesize{\small}
\tablecaption{\label{tab:tHODz2}The initial and final HOD parameters for galaxies passively flowing from $z=2$}
\startdata\hline\hline
Model&$\al$ & $\Ml/\Mm$ &$\Mlft$\tablenotemark{a}&$\Mmft$\tablenotemark{a}&$\fcen$&$\al$ & $\Ml/\Mm$&$\Mlft$\tablenotemark{a}&$\Mmft$\tablenotemark{a}&$\fcen$\\ 
    &           && $z=2$  &       &      &              & & $z=0.3$&     & \\\hline
21  &0.5& 1.986&0.425 &0.2140 &0.60  &0.927& 1.754&1.601  &0.9131 &0.50\\  
22  &1& 2.120  &0.425 &0.2005 &0.67  &1.11&2.226  &1.877  &0.8431 &0.55\\
23  &2& 2.144  &0.425 &0.1982 &0.68  &1.46& 2.454 &2.040  &0.8311 &0.56\\ 
24  &0.5& 9.167&1.7   &0.1855 &0.76  &1.11& 3.162 &2.422  &0.7660 &0.62\\ 
25  &1& 10.23  &1.7   &0.1661 &0.90  &1.15& 4.151 &2.783  &0.6705 &0.72\\
26  &2& 10.78  &1.7   &0.1577 &0.98  &1.20& 4.814 &3.022  &0.6277 &0.78\\
27  &0.5& 24.34&4.25  &0.1746 &0.83  &1.131& 3.762&2.679  &0.7121 &0.67\\
28  &1& 26.60  &4.25  &0.1598 &0.96  &1.17& 4.651 &2.978  &0.6403 &0.76\\
29  &2& 27.30  &4.25  &0.1557 &0.995 &1.18& 5.134 &3.173  &0.6180 &0.79\\
30  & &        &      &0.1553 &1     &1.18& 5.234 &3.225  &0.6162 &0.79\\
31  & &        &      &0.1056 &1     &1.12& 4.679 &3.735  &0.7983 &0.80\\
\enddata
\tablenotetext{a}{Mass of halos are in the unit of $10^{14} \Msun$.}
\tablecomments{We show the input HOD parameters at $z=2$ and the best fit HOD parameters at $z=0.3$ for \ztM.  Note that $\Ml$ and $\al$ at $z=0.3$ are fitted over $M>\Ml$ in order to better describe the shape of satellite HODs at the massive end. The value $\fcen$ is the fraction of central galaxies to the total number of galaxies. }
\end{deluxetable}

\begin{deluxetable}{l|ccccc|ccccc}
\tablewidth{0pt}
\tabletypesize{\small}
\tablecaption{\label{tab:tHODz1n4} HOD parameters at $z=1$ with four times the number density of the LRGs}
\startdata\hline\hline
Model&$\al$ & $\Ml/\Mm$&$\Mlft$\tablenotemark{a}&$\Mmft$\tablenotemark{a}&$\fcen$&$\al$ & $\Ml/\Mm$&$\Mlft$\tablenotemark{a}&$\Mmft$\tablenotemark{a}&$\fcen$\\ 
    &           & & $z=1$  &       &      &            & & $z=0.3$&     & \\\hline
7\nfo &0.5& 19.70&2.6  &0.1320 &0.78  &0.927& 5.502&1.175  &0.2135 &0.656\\
8\nfo &1& 22.61  &2.6  &0.1150 &0.907 &0.995& 7.574&1.403  &0.1852 &0.83\\
9\nfo &2& 23.91  &2.6  &0.1088 &0.97  &1.132& 9.609&1.681  &0.1749 &0.83\\
10\nfo& &        &     &0.1062 &1.0   &0.952& 10.27&1.7532 &0.1707 &0.83\\
\enddata
\tablenotetext{a}{Mass of halos are in the unit of $10^{14} \Msun$.}
\tablecomments{We show the input HOD parameters at $z=2$ and the best fit HOD parameters at $z=0.3$ for \zoMfn.  Note that $\Ml$ and $\al$ at $z=0.3$ are fitted over $M>\Ml$ in order to better describe the shape of satellite HODs at the massive end. The value $\fcen$ is the fraction of central galaxies to the total number of galaxies. }
\end{deluxetable}

\begin{deluxetable}{l|cccc}
\tablewidth{0pt}
\tabletypesize{\small}
\tablecaption{\label{tab:tHODz0.3} The five-parameter HOD at $z=0.3$ for Model LRG}
\startdata\hline\hline
Model&$\al$ & $\Mlft'$\tablenotemark{a}&$\Mmft'$\tablenotemark{a}&$\fcen$ \\ \hline 
LRG  &1.86 (1.84) & 6.875 (6.875) & 0.8226 (0.7764) &0.931 (0.932)\\
\enddata
\tablenotetext{a}{Mass of halos are in the unit of $10^{14} \Msun$.}
\tablenotetext{b}{We use $M_0=3.209$ (3.056)$\times 10^{9} \Msun$ and $\sigma_M=0.556$ (0.556).}
\tablenotetext{c}{Numbers in the parentheses are the best fit values by \citet{Zheng07} for our fiducial cosmology.}
\tablenotetext{d}{The number density of galaxies for Model LRG is $9.817\times 10^{-5} \itrihMpc$. This is only slightly different from the fiducial number density $10^{-4} \itrihMpc$ that we adopted for passive flow.}
\tablecomments{HOD parameters at $z=0.3$ in our simulations that correspond to the best fit HOD by \citet{Zheng07} for the observed LRG clustering. For comparison, when Model LRG is fitted to the three-parameter HOD (eq. [\ref{eq:Ngal}]), we find $\al=1.86$, $\Ml/\Mm=12.2$, $\Ml=6.790\times 10^{14}\Msun$, and $\Mm=0.5563\times 10^{13}\Msun$.}
\end{deluxetable}

\end{document}